\documentclass[12pt,nofootinbib,preprint]{revtex4}
\pdfoutput=1

\usepackage{amsmath}
\usepackage{amssymb}
\usepackage{anysize}
\usepackage{bold-extra}
\usepackage{xcolor}
\usepackage{enumerate}
\usepackage{fancyhdr}
\usepackage{graphicx}
\usepackage[linktocpage,colorlinks,urlcolor=blue,citecolor=blue,linkcolor=blue]{hyperref}
\usepackage{multirow}
\usepackage[final]{pdfpages}
\usepackage{rotating}
\usepackage{textcomp}
\usepackage{url}
\usepackage{verbatim}
\usepackage{wrapfig}
\usepackage{slashed}
\usepackage{subfigure}

\usepackage{amsfonts}
\usepackage{comment}
\usepackage{epigraph}
\usepackage[utf8]{inputenc}
\usepackage{slashed}
\usepackage{bbm}
\usepackage{cancel}
\usepackage{epsf, array, color}
\newcommand{\La}{\mathcal{L}}

\newcommand{\sla}[1]{/\!\!\!\!#1}

\begin{document}

\title{\boldmath QCD Corrections in SMEFT Fits to $WZ$ and $WW$ Production}

\author{Julien Baglio$^{1}$}
\email{julien.baglio@cern.ch}

\author{Sally Dawson$^{2}$}
\email{dawson@bnl.gov}

\author{Samuel Homiller$^{2,3}$}
\email{samuel.homiller@stonybrook.edu}

\affiliation{$^{1}$Institute for Theoretical Physics, University of
  T\"ubingen, Auf der Morgenstelle 14, 72076 T\"ubingen, Germany\\
$^{2}$Department of Physics, Brookhaven National Laboratory, Upton, N.Y. 11973~ U.S.A.\\
$^{3}$C. N. Yang Institute for Theoretical Physics, Stony Brook University, 
 Stony Brook, NY 11794~ U.S.A.
 }

\begin{abstract}
  We investigate the role of anomalous gauge boson and fermion
  couplings on the production of $WZ$ and $W^+W^-$  pairs at the  LHC
  to NLO QCD in the Standard Model effective field theory, including
  dimension-6 operators. Our results are implemented in a publicly
  available version of the {\tt POWHEG-BOX}. We combine our $WZ$
  results in the leptonic final state $e\nu \mu^+\mu^-$ with previous
  $W^+W^-$ results to demonstrate the numerical effects of NLO QCD
  corrections on the limits on effective couplings derived from ATLAS
  and CMS $8$ and $13$\,TeV  differential measurements.
  Our study demonstrates the importance of including NLO QCD SMEFT
  corrections in the $WZ$ analysis, while the effects on $WW$
  production are smaller. We also show that the
  $\mathcal{O}(1/\Lambda^4)$  contributions dominate the analysis, where $\Lambda$ is
  the high energy scale associated with the SMEFT.

\end{abstract}

\preprint{YITP-SB-19-28}

\maketitle

\section{Introduction}

The properties of the Standard Model (SM) have been experimentally
verified at the LHC at the ${\cal{O}}(10-20\%)$ level in the Higgs
sector~\cite{Dawson:2018dcd} and there is no evidence for the
existence of any new particles or interactions at the TeV scale yet.
High statistics measurements of gauge boson pair production allow  for
detailed comparisons with Standard Model predictions and  can be used
to quantify the restrictions on anomalous interactions. Gauge boson
pair production is particularly sensitive to new $3$-gauge boson
interactions~\cite{Hagiwara:1986vm} or new fermion-boson
interactions~\cite{Zhang:2016zsp}. The current task is to make
comparisons between theory and data at the few percent level which
requires not only high-luminosity LHC running, but also improved
theoretical calculations.

The SM rates for both $W^+W^-$ and $WZ$ production are well known. QCD
corrections to $WZ$  production in the Standard Model  have been
computed to next-to-leading order (NLO) for on-shell
production~\cite{PhysRevD.44.3477, Frixione:1992pj} and to NNLO for
both on- and off-shell
production~\cite{Grazzini:2016swo,Grazzini:2017ckn}. SM electroweak
corrections to the $WZ$
process~\cite{Bierweiler:2013dja,Baglio:2013toa,Biedermann:2017oae,Baglio:2018rcu,Denner:2019tmn}
are also known at NLO and can have significant effects in the high
$p_T$ regime. $W^+W^-$ pair production is also under good theoretical
control in the SM: NNLO
QCD~\cite{Gehrmann:2014fva,Caola:2015rqy,Grazzini:2016ctr} and NLO
electroweak~\cite{Baglio:2013toa,Bierweiler:2013dja,Biedermann:2016guo}
corrections are understood and change the distributions and rates
significantly.

Gauge boson pair production can be put under the microscope using  an
effective Lagrangian,
\begin{equation}
L_{\rm SMEFT}=
 L_{\rm SM}+\sum_{i,n}{C_i^{(n)}\over\Lambda^{n-4}}O_i^{(n)}+\ldots\, ,
 \label{eq:smeft}
\end{equation}
where
the new physics is parameterized as an operator expansion in inverse
powers of a high scale $\Lambda$ and the assumption is made that there
are no light degrees of freedom. The operators $O_i^{(n)}$ have mass
dimension-$n$, are invariant under $SU(3)\times SU(2)\times U(1)$ and
$L_{\rm SM}$ contains the complete SM Lagrangian.  The subscript SMEFT
indicates that the Higgs is taken to be part of an $SU(2)$ doublet.
At dimension-6, there are 59 possible
operators~\cite{Buchmuller:1985jz,Grzadkowski:2010es} when flavor
effects are neglected. We compute the amplitudes  for $W^+W^-$ and
$WZ$ pair production including the dimension-6 operators, and then
consider results when the cross sections are consistently expanded to
both $1/\Lambda^2$  and $1/\Lambda^4$.

The leptonic decay channel of $W^+W^-$ pair production has been
studied  at NLO in the SMEFT in a previous
work~\cite{Baglio:2018bkm}. Here we extend those results to include
the leptonic decays from  $WZ$ pair production at the LHC in the
presence of anomalous 3- gauge boson and  anomalous fermion- gauge
boson couplings. QCD effects can affect the dependence of the
kinematic distributions on the coefficients of Eq.~(\ref{eq:smeft}). 
We include  anomalous 3-gauge boson couplings and anomalous
fermion-gauge boson couplings in the {\tt POWHEG-BOX} to NLO QCD in
the SMEFT
approach~\cite{Dixon:1998py,Dixon:1999di,Baur:1994aj,Azatov:2019xxn}
following previous implementations for the SMEFT 3-gauge boson
couplings case~\cite{Melia:2011tj,Nason:2013ydw}. This public tool can
be found  at  \url{http://powhegbox.mib.infn.it}.

Limits on SMEFT coefficients  have been obtained in global fits that
include  gauge boson pair production, Higgs measurements, electroweak
precision measurements, and top quark
measurements~\cite{deBlas:2017wmn,Ellis:2018gqa,Grojean:2018dqj,Almeida:2018cld,Biekotter:2018rhp}. The
SMEFT effects are treated at tree level in these fits, while the SM
results include all known higher order SM predictions. Fits attempting
to use full NLO electroweak SMEFT predictions quickly observe that the
plethora of operators makes such fits
problematic~\cite{Berthier:2016tkq,Dawson:2019clf}. On the other
hand, the inclusion of NLO QCD SMEFT effects is simpler, due to the
smaller number of operators
involved~\cite{Baglio:2018bkm,Baglio:2017bfe,Hartland:2019bjb}.

In Section~\ref{sec:basics}, we define our notation in terms of
anomalous couplings and present some calculational
details. Section~\ref{sec:fp} contains  a sampling of kinematic
distributions  with benchmark values of the anomalous couplings and
Section \ref{sec:fits} has the results of a numerical fit to $W^+W^-$
and $WZ$ data. The NLO SMEFT QCD corrections have a numerically
significant effect on many of the results. We point out that fits to
${\cal{O}(}{1/\Lambda^2})$ or to ${\cal{O}}({1/\Lambda^4})$ result in
quite different limits on the SMEFT coefficients. We conclude in
Section \ref{sec:conc}. We also provide fits using
$W^+W^-$ data only in Appendix~\ref{app:ww_limits}, while a discussion
about the truncation at order $\mathcal{O}(1/\Lambda^2)$ is carried in
Appendix~\ref{app:truncation}.

\section{Basics}
\label{sec:basics}

\subsection{Effective Gauge and Fermion Interactions}
\label{sec:eftres}

We begin by reviewing the most general CP and Lorentz invariant Lagrangian 
for anomalous $W^+W^-Z$ and $W^+W^-\gamma$ couplings ~\cite{Gaemers:1978hg,Hagiwara:1986vm},
\begin{eqnarray}
 \La_{V}=
-ig_{WWV}\biggl[g_1^V\left(W^+_{\mu\nu}W^{-\mu}V^\nu-W_{\mu\nu}^-W^{+\mu}V^\nu\right)+\kappa^VW^+_\mu
            W^-_\nu V^{\mu\nu}+\frac{\lambda^V}{M^2_W}W^+_{\rho\mu}{W^{-\mu}}_\nu V^{\nu\rho}\biggr],
\label{eq:lagdef}
\end{eqnarray}  
with $V=\gamma, Z$, $g_{WW\gamma}=e$, $g_{WWZ}=g \cos\theta_W$,
($s_W^{} \equiv \sin\theta_W^{}$, $c_W^{} \equiv
\cos\theta_W^{}$).  
The anomalous couplings are defined as $g_1^V = 1+\delta g_1^V$, $\kappa_{}^V=
1+\delta\kappa_{}^V$, where in the SM  $\delta g_1^V = \delta\kappa_{}^V
= \lambda_{}^V = 0$ and gauge invariance implies
$\delta g_1^\gamma = 0$.

The effective couplings of quarks to gauge fields are \footnote {We assume
no new tensor structures and neglect CKM mixing and all flavor effects. We assume SM gauge couplings to leptons, since these couplings
are highly restricted by LEP data.  We further  neglect possible anomalous right-handed $W$-quark couplings, since they
are suppressed by small Yukawa couplings in an MFV framework and stringently limited by Tevatron and LHC measurements }~\cite{Baglio:2018bkm,Falkowski:2014tna,Berthier:2016tkq,Zhang:2016zsp}, 
\begin{eqnarray}
  \La&\equiv &g_ZZ_\mu\biggl[g_L^{Zq}+\delta g_{L}^{Zq}\biggr]
  {\overline q}_L\gamma_\mu q_L\
 +g_ZZ_\mu\biggl[g_R^{Zq}+\delta g_{R}^{Zq}\biggr]
  {\overline q}_R\gamma_\mu q_R\nonumber \\
  &&+{g\over \sqrt{2}}\biggl\{W_\mu\biggl[(1+\delta g_{L}^W){\overline u}_L\gamma_\mu d_L
  +\delta g_R^W
  {\overline u}_R\gamma_\mu d_R\biggr] +h.c.\biggr\}\, .
  \label{eq:dgdef}
  \end{eqnarray}
Here, $g_Z=e/(c_W^{}s_W^{})= g/c_W$ and $q$ is an  up- or down-flavor quark.  The SM quark interactions  are:
\begin{eqnarray}
g_R^{Zq}&=&-s_W^2 Q_q\quad{\rm and}\quad g_L^{Zq}=T_3^q -s_W^2 Q_q,
\end{eqnarray}
where $T_3^q=\pm \displaystyle \frac{1}{2}$ and $Q_q$ is the electric
 charge.

 $SU(2)$ invariance implies,
\begin{eqnarray}
\delta g_L^W&=&\delta g_L^{Zu}-\delta g_L^{Zd},
\nonumber \\
\delta g_1^Z&=& \delta \kappa_{}^Z+{s_W^2\over c_W^2}\delta \kappa_{}^\gamma,
\nonumber \\
\lambda_{}^\gamma &=& \lambda_{}^Z\, .
\label{eq:su2rel}
\end{eqnarray}
This framework leads
to $7$ unknown parameters, $\delta g_1^Z,~ \delta \kappa_Z,~\lambda_Z, ~\delta g_L^{Zu},~\delta g_L^{Zd},~\delta g_R^{Zu}$
and $~\delta g_R^{Zd}$,  contributing to $W^+W^-$ production.  The anomalous right-handed couplings do not contribute to $WZ$
production, hence reducing the number of unknown parameters down to $5$.  These parameters are ${\cal{O}}({1/ \Lambda^2})$ in the SMEFT language.
The conversion between the effective Lagrangians of Eqs. \ref{eq:lagdef} and \ref{eq:dgdef}   and the
dimension-6 interactions in the Warsaw basis can be found in many places~\cite{Baglio:2018bkm,Zhang:2016zsp,Berthier:2015oma}
and there is a one-to-one mapping between the two approaches\footnote{See for example, Tables 4 and 5 of Ref.~\cite{Baglio:2018bkm}.}.

It is of interest to study the high energy limits of the helicity amplitudes for $W^+W^-$ and $WZ$ scattering in order to understand
generic features of our results. 
In the high energy limit ($s\gg M_Z^2$), only the longitudinal $(00)$ and  transverse $(\pm \mp)$ 
helicity amplitudes remain non-zero in the SM $WZ$
amplitudes
(where $s$ is the partonic center of mass energy-squared)~\cite{Baur:1994ia},
\begin{eqnarray}
A_{00}^{SM,W^+Z} &\rightarrow & -{g^2  \over 2\sqrt{2}}\sin\theta
\nonumber \\
 A_{\pm,\mp}^{SM,W^+Z}
  &\rightarrow&{g^2\over \sqrt{2}}c_W\biggl({1-\cos\theta\over \sin\theta}\biggr)\biggl[\cos\theta+{1\over 3}\tan^2\theta_W\biggr]
\nonumber \\
 A_{\pm,\pm}^{SM,W^+Z}
  &\rightarrow&
  {\cal{O}}\biggl({M_Z^2\over s}\biggr)
   \, ,
 \label{eq:wzamps}
 \end{eqnarray}
where $\theta$ is the center of mass  angle of the W boson with respect to the up quark direction and $g_L^{Zu}-g_L^{Zd}=c_W^2$. 
The radiation zero  in the high energy  $(\pm,\mp)$ amplitude at $\cos\theta_0=(g_L^{Zu}+g_L^{Zd})/(g_L^{Zu}-g_L^{Zd})$ is clearly seen in Eq. \ref{eq:wzamps}. 

The SMEFT contributions  to $W^+Z$ production  that contribute interference effects with the SM in the high 
energy limit are~\cite{Baur:1994ia,Grojean:2018dqj},
\begin{eqnarray}
\delta A_{00}^{W^+Z}  &\rightarrow& 
{g^2\over 2\sqrt{2} } 
\sin\theta 
\biggl({s\over M_Z^2}\biggr)\biggl[\delta g_1^Z+{(\delta g_L^{Zd}-\delta g_L^{Zu})\over c_W^2}
\biggr]
\nonumber \\ 
\delta A_{\pm\pm}^{W^+Z}&\rightarrow &
{g^2\over 2\sqrt{2} c_W}\sin\theta  \biggl({s\over M_Z^2}\biggr)\lambda_Z
\nonumber \\
\delta A_{\pm,\mp}^{W^+Z}&\rightarrow &
-{g^2\over \sqrt{2} c_W}\sin\theta \biggl[ \delta g_L^{Zu}\tan^2({\theta\over 2})+\delta g_L^{Zd}\biggr]
\,.
\label{eq:wzlims}
\end{eqnarray}
Note that in the high energy limit, $s\gg M_Z^2$, the dependence on $\delta \kappa_Z$ is suppressed and that the energy enhanced
longitudinal amplitude peaks at $\theta={\pi\over 2}$.  Only the longitudinal modes have an  energy enhanced interference
contribution with the SM. The approximate zero of the  SM $(\pm\mp)$ amplitude is weakened in the high energy
limit where contributions from the anomalous fermion couplings fill in the dip at $\cos\theta_0$.

The complete helicity amplitudes for $W^+W^-$ production can be found in~\cite{Hagiwara:1986vm,Baglio:2017bfe}.   
 The energy enhanced amplitudes for $q_{L,R} {\overline {q}}_{L,R}\rightarrow W^+W^- $are,
\begin{eqnarray}
\delta A_{LL00}^{W^+W^-}&=& {g^2\over 2}{s\over M_W^2}\sin\theta\biggl[
 \delta \kappa_Z(Q_q -T_3^q)-c_W^2 Q_q \delta g_1^Z-\delta g_L^{Zq}
+2T_3^q\delta g_L^W\biggr]
\nonumber \\
\delta A_{RR00}^{W^+W^-}&=& {g^2\over 2}{s\over M_W^2}\sin\theta\biggl[
 -Q_q\delta \kappa_Z+c_W^2 Q_q \delta g_1^Z+\delta g_R^{Zq}
\biggr]\, .
\label{eq:wwlims}
\end{eqnarray}
Due to the Goldstone boson nature of the longitudinal modes, the amplitudes of Eqs. \ref{eq:wzlims}
and \ref{eq:wwlims} satisfy
\begin{equation}
 \delta A_{00}^{W^+Z}  =\delta A_{LL00}^{W^+W^-}({\overline u}_Lu_L\rightarrow W^+W^-) 
  -\delta A_{LL00}^{W^+W^-}({\overline d}_Ld_L\rightarrow W^+W^-)  
 \, .
\end{equation}
This implies that the high energy limits of $WV$ production (V=W,Z) are only sensitive to 4 combinations of coefficients, and the dependence on other parameters is suppressed by powers of ${M_W^2\over s}$~\cite{Gupta:2014rxa,Falkowski:2016cxu,Grojean:2018dqj,Franceschini:2017xkh,}.

The amplitudes for  $W^+W^-$ and $WZ$ production can be schematically written as,
\begin{equation}
A\sim A_{SM}+{\delta A_{EFT}^{(6)}\over \Lambda^2}+
{\delta A_{EFT}^{(8)}\over \Lambda^4}+...
\, .
\end{equation}
The cross section is then,
\begin{equation}
\sigma\sim {1\over s}\biggl[\mid A_{SM}\mid^2+
2\,\mathrm{Re}\biggl({A_{SM}^*\delta A_{EFT}^{(6)}\over\Lambda^2}\biggr)+{\mid \delta A_{EFT}^{(6)}\mid^2\over \Lambda^4}
+2\,\mathrm{Re}\biggl({A_{SM}^*\delta A_{EFT}^{(8)}\over\Lambda^4}\biggr)+.... \biggr] 
\, .
\end{equation}
In the results of this paper, we implicitly assume that the
${1/\Lambda^4}$ contributions from the dimension-8 operators can be
neglected, so that terms of ${\cal {O}}({1/ \Lambda^4})$
in the cross section can be kept in a consistent way. This is the
case, for example, in strongly interacting
models~\cite{Giudice:2007fh,Contino:2016jqw}.
We present an analysis of the truncation at ${\cal
{O}}({1/ \Lambda^2})$ in the Appendix~\ref{app:truncation}, to check
whether or not the ${\cal {O}}({1/ \Lambda^4})$ terms are important.

\subsection{Primitive Cross Sections}
\label{sec:prim}
We want to compute differential and total cross sections for the $WZ$ scattering process at NLO QCD for arbitrary
anomalous couplings with kinematic cuts mimicing the experimental analyses. The current calculation  uses identical techniques as in Ref.~\cite{Baglio:2018bkm}.
The decomposition into primitive cross sections  works at both lowest order (LO) and NLO and there are $15$ primitive cross sections for the $WZ$ process and $35$ for the $W^+W^-$ process at ${\cal{O}}(\Lambda^{-4})$.   

\subsection{Calculational Details} 
We have implemented the process $pp\rightarrow WZ\rightarrow( l^+l^-)( l^\prime \overline {\nu}^\prime )$ into the {\tt POWHEG-BOX-V2} including
anomalous fermion and gauge boson couplings. The existing implementation~\cite{Melia:2011tj} does not allow for anomalous fermion couplings. 
Our new implementation allows the user to chose the order of the $\Lambda^{-2n}$ expansion
and to use either the effective Lagrangians described in this work or the Warsaw basis coefficients.  Note that we assume different flavor 
leptonic decays.  The results shown in the following sections use  CTEQ14qed  PDFs and we 
fix the renormalization/factorization scales at $M_Z/2$. 
 
\section{NLO Effects in $WZ$ Distributions}
\label{sec:fp}

We now present distributions for various kinematic variables at LO and NLO with different values of the anomalous couplings using the methods described in the previous section.
In addition to the Standard Model, we present results for two benchmark points:
\begin{eqnarray}
\mathrm{Gauge~(or~3GB):} & \quad  \delta g_1^Z = 0.016,~ \lambda^Z = 0.0045,~ \delta \kappa^Z = 0.024; \quad \delta g_L^{Zu} = 
\delta g_L^{Zd} = 0\nonumber \\
\mathrm{Fermion~(or~Ferm.): } & \quad \delta g_1^Z = \lambda^Z = \delta \kappa^Z = 0; \quad \delta g_L^{Zu} = -0.0024,~ \delta g_L^{Zd} = 0.003
\, .
\label{eq:bench}
\end{eqnarray}
Both of these points are near the boundaries of the allowed regions from fits to $W^+W^-$ and $WZ$ production 
and serve to illustrate the effects of anomalous couplings on the NLO QCD corrections.

\begin{figure}
\centering
\includegraphics[width=0.49\linewidth]{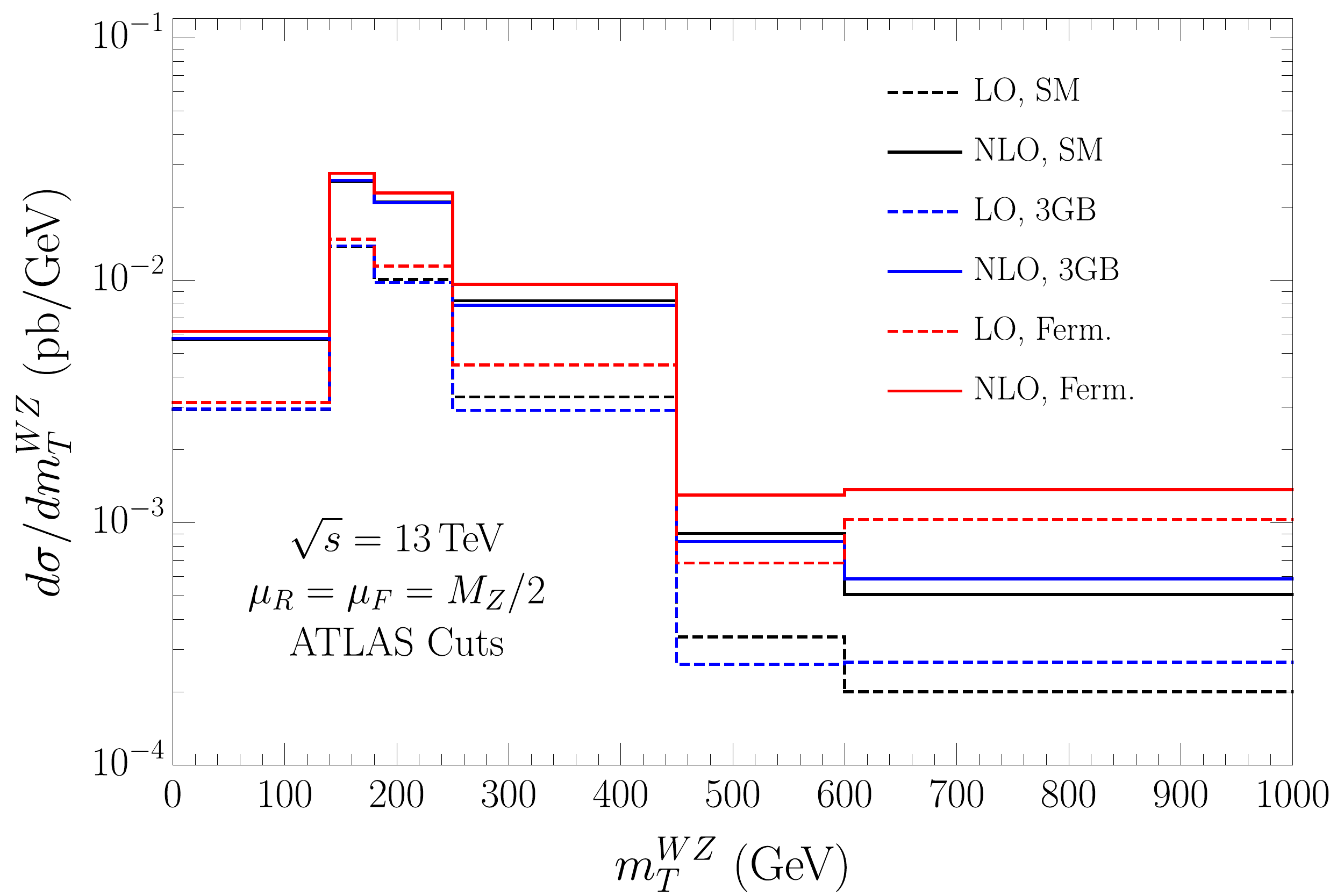}
\includegraphics[width=0.49\linewidth]{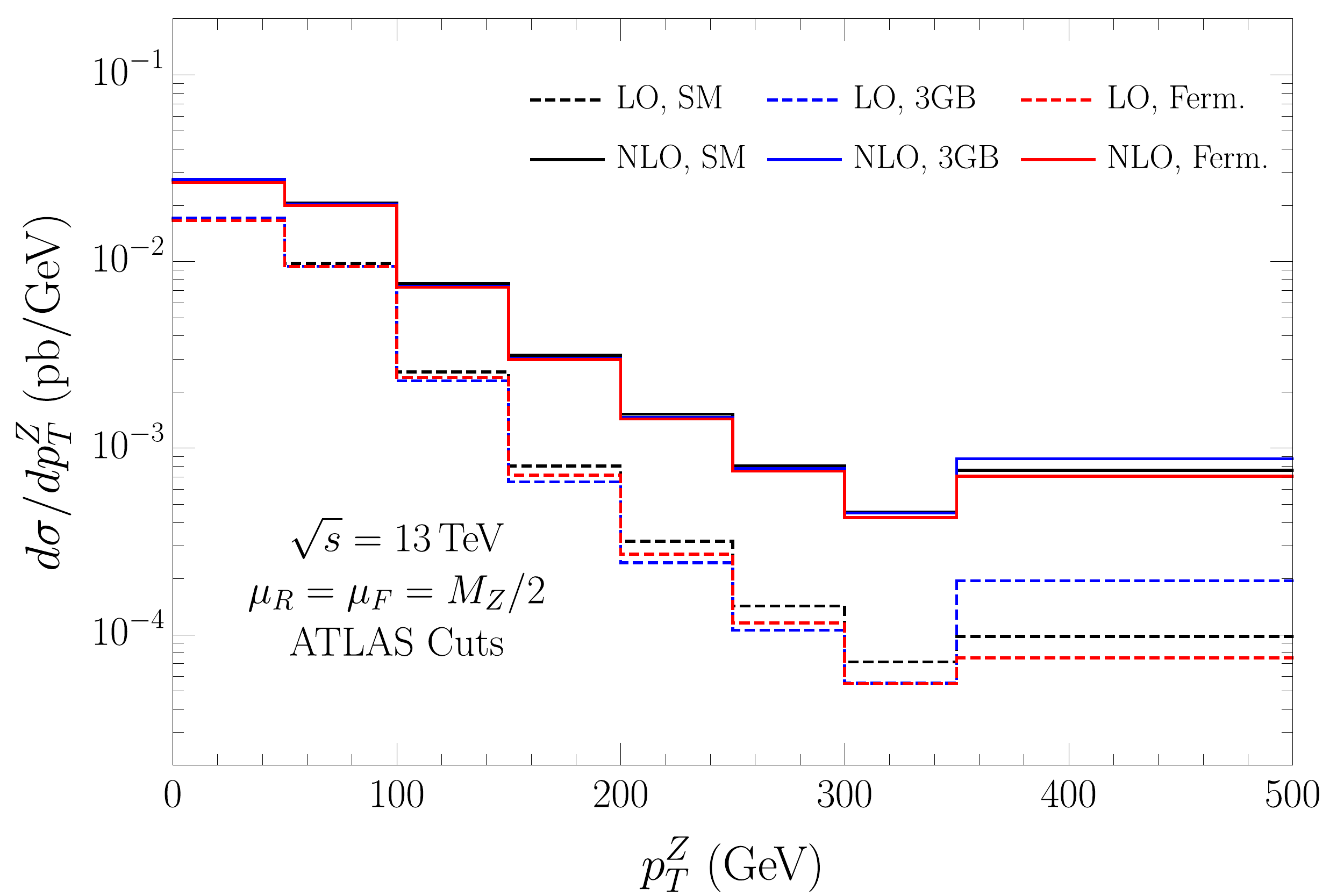}\\
\vspace{0.2cm}
\includegraphics[width=0.49\linewidth]{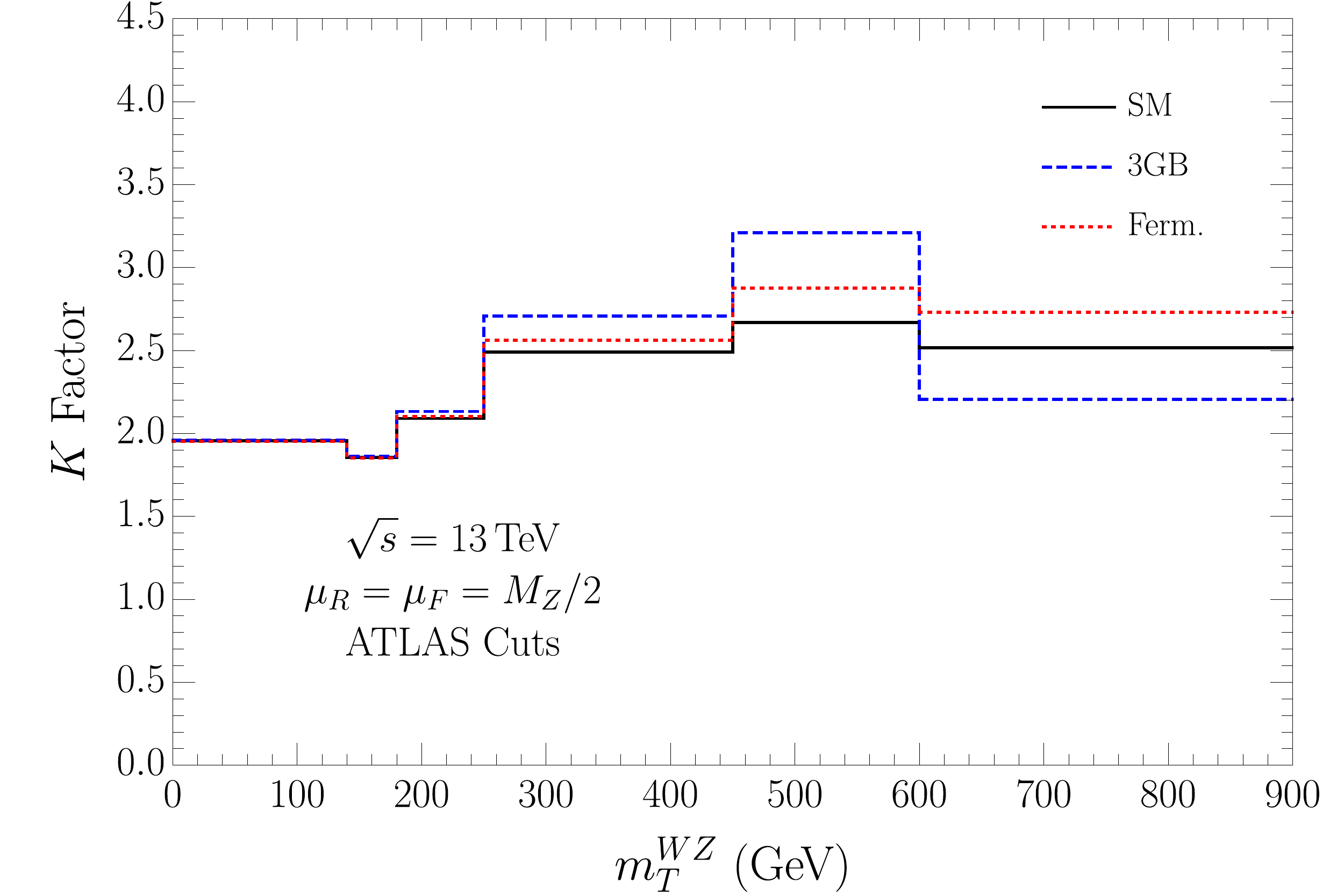}
\includegraphics[width=0.49\linewidth]{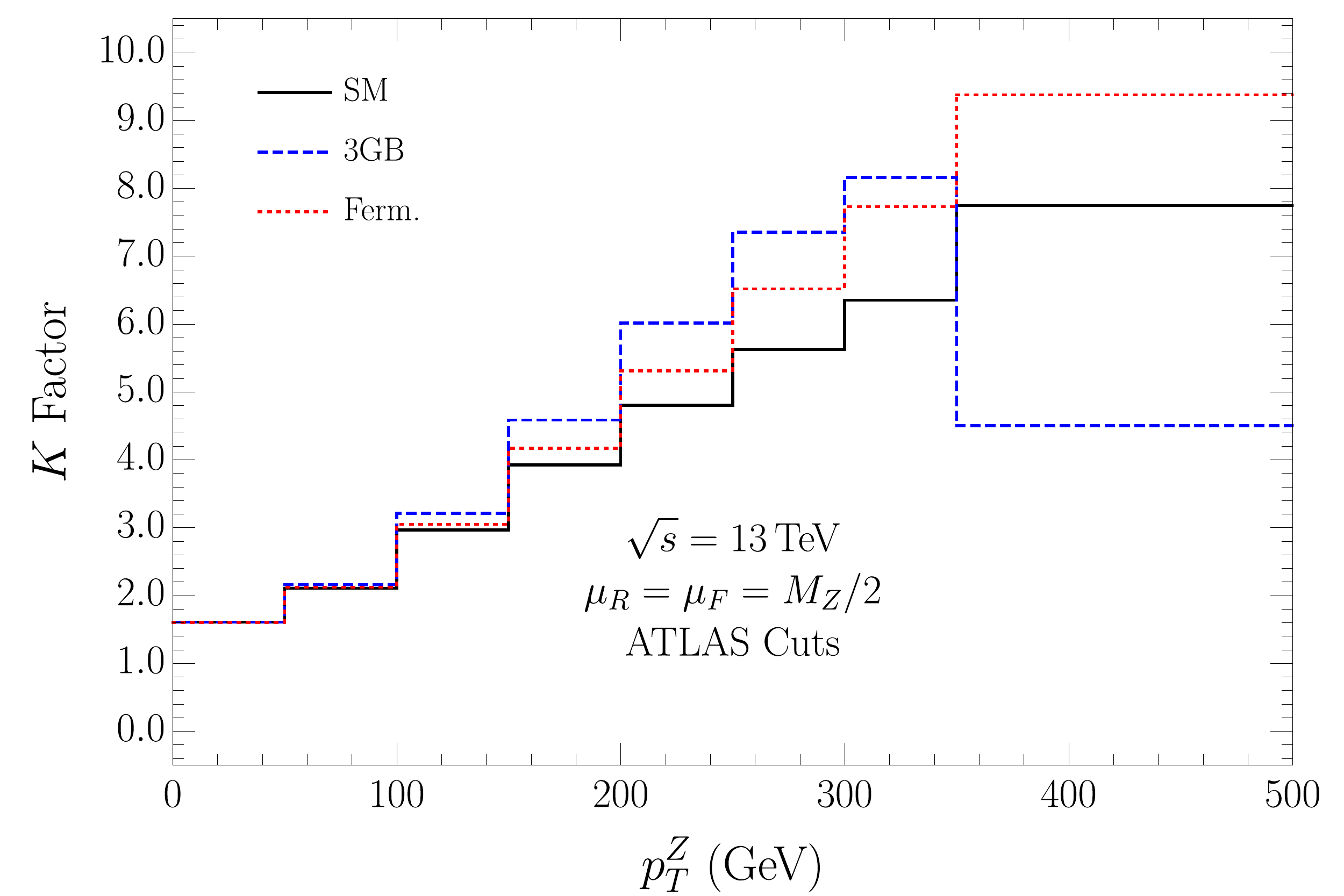}
\caption{Top Row: Distributions at LO and NLO for the SM, anomalous gauge benchmark point and 
anomalous fermion benchmark  point, Eq. \ref{eq:bench},  in bins of $m_T^{WZ}$ and $p_T^{Z}$.\\
Bottom Row: K-factors for the same three points. In both the $m_T^{WZ}$ and $p_T^Z$ distributions the final bin goes  to $2\,\mathrm{TeV}$.
}\label{fig:dists}
\end{figure}

In Fig.~\ref{fig:dists}, we show the distributions in bins of $m_T^{WZ}$ and $p_T^Z$ 
along with the corresponding ratios of the NLO and LO predictions using the cuts from Ref.~\cite{Aaboud:2019gxl}, where
\begin{equation}
m_T^{WZ} = \sqrt{
\left(\sum_{\ell=1}^{3} p_{T}^{\ell} + E_T^{\mathrm{miss}}\right)^2 - 
\left(\sum_{\ell=1}^{3} p_{x}^{\ell} + E_x^{\mathrm{miss}}\right)^2 - 
\left(\sum_{\ell=1}^{3} p_{y}^{\ell} + E_y^{\mathrm{miss}}\right)^2
}\,  .
\end{equation}

In the right panel we see that at high $p_{T,Z}$ the $K$~factor\footnote{The $K$ factor is defined as the ratio of the NLO/LO result
for a given scenario.}  for the SM becomes very large as a result of real emission effects that arise at NLO, in agreement with 
Refs.~\cite{Denner:2019tmn,Baglio:2018bkm}. 
In contrast, the $K$~factor grows only modestly as a function of $m_T^{WZ}$.
For the anomalous coupling benchmarks, we see that the $K$~factor can change quite dramatically, particularly
 in the higher-momentum bins~\cite{Baur:1994aj}. In the last $p_T^Z$ bin in particular, the $K$~factor changes from $\sim 7.8$ in the SM to roughly $4.5$ for our ``Gauge" benchmark, and $\sim 9.4$ for the ``Fermion" point. Similar, but less dramatic, effects are seen in $m_T^{WZ}$ as well. 

The results of Fig.~\ref{fig:dists} clearly demonstrate that using the Standard Model $K$~factor in an analysis of anomalous couplings in $WZ$ production is inaccurate at large $p_T^Z$. As the high transverse momentum  bins provide most of the constraining power for fits to the anomalous couplings, this can drastically change the resulting limits on the anomalous coefficients, as we demonstrate in the following sections.

\begin{figure}
\centering
\includegraphics[width=0.49\linewidth]{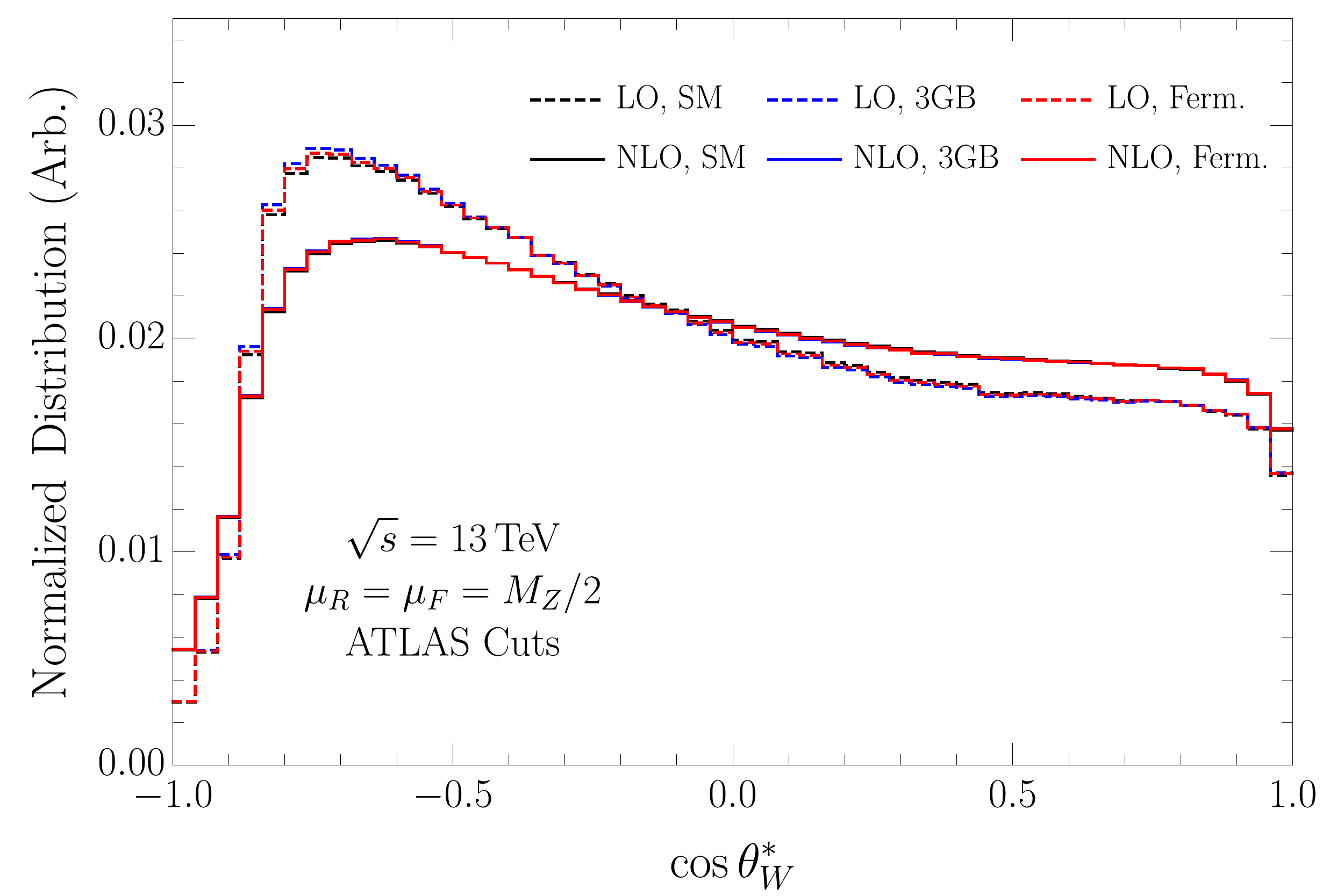}
\includegraphics[width=0.49\linewidth]{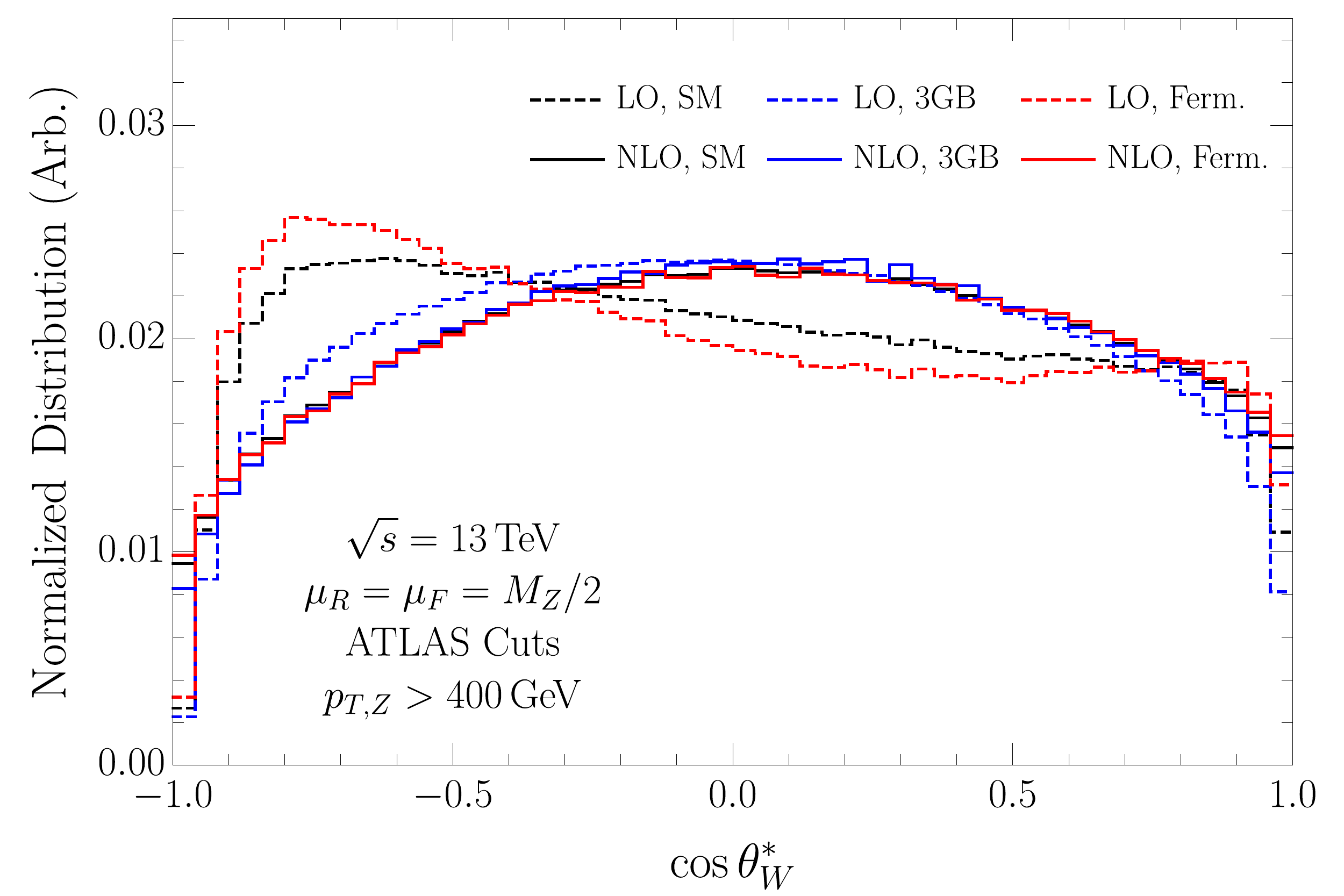}\\
\vspace{0.2cm}
\includegraphics[width=0.49\linewidth]{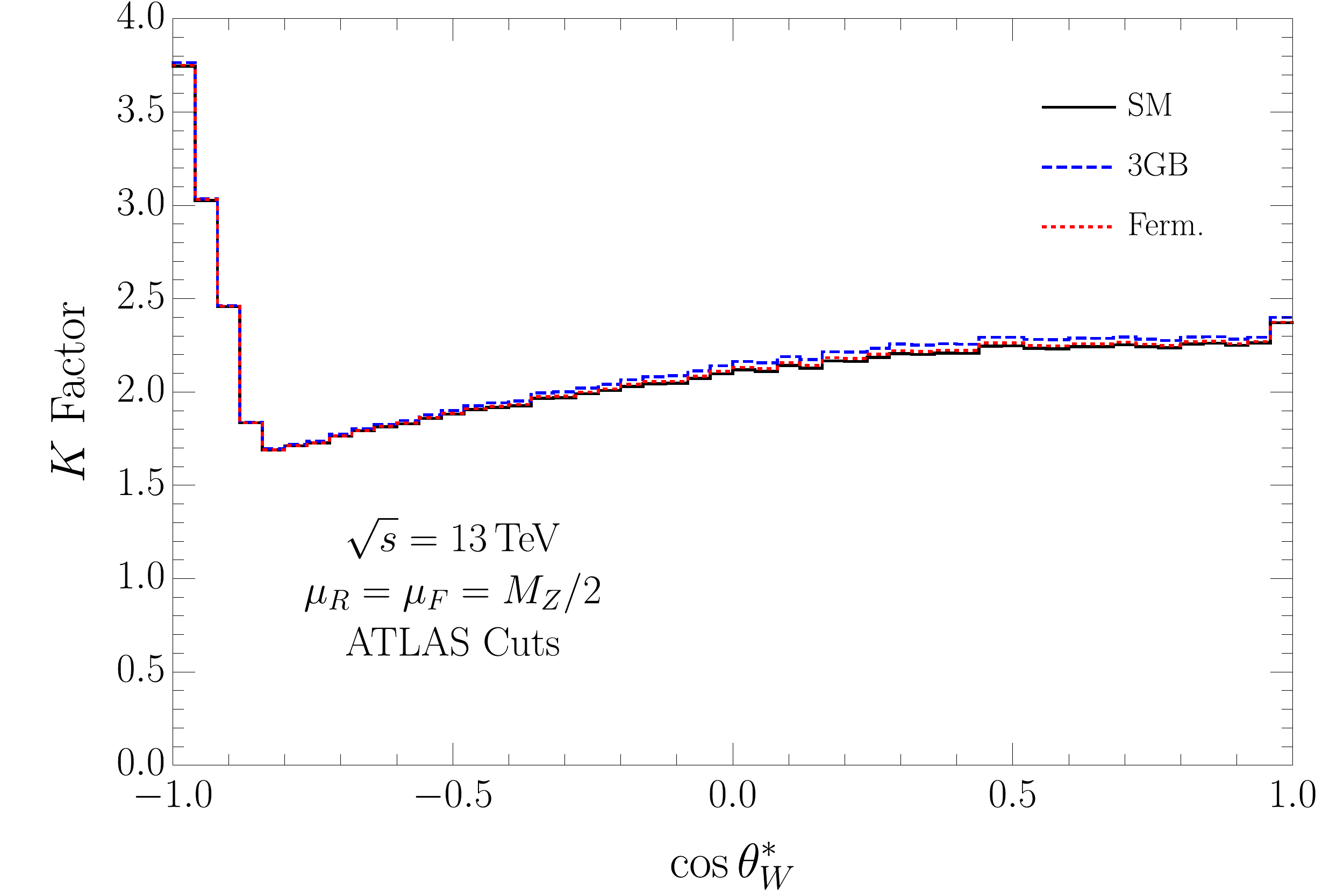}
\includegraphics[width=0.49\linewidth]{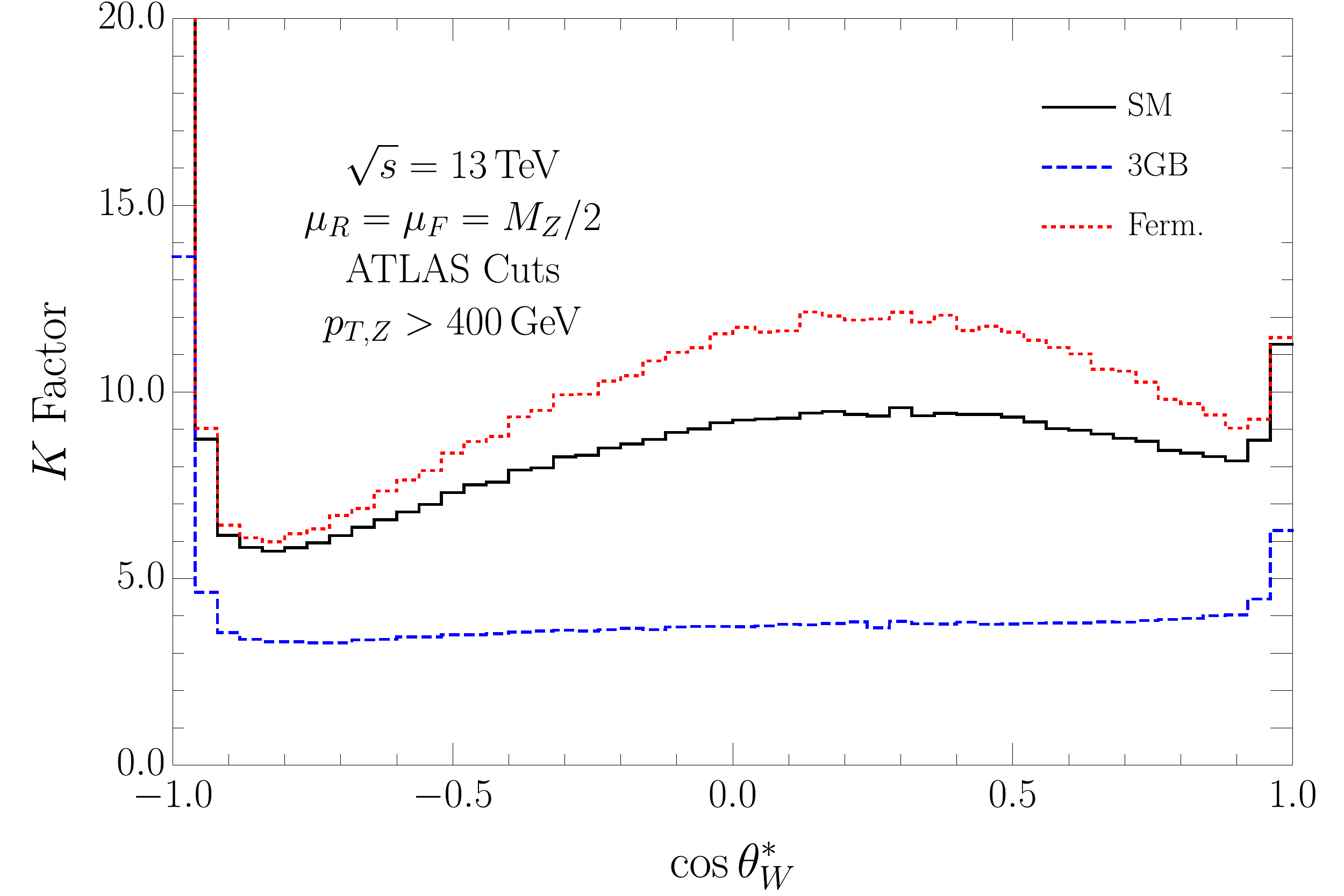}
\caption{
Top Row: Normalized distribution at LO and NLO for the SM, anomalous gauge interaction point and anomalous fermion benchmark points,
Eq. \ref{eq:bench},  in bins of $\cos\theta_W^*$ with the ATLAS $13\,\mathrm{TeV}$ fiducial cuts~\cite{Aaboud:2019gxl}. The left panel shows the inclusive distribution,  while the right panel shows the distribution after an additional cut requiring $p_{T,Z} > 400\,\mathrm{GeV}$.
Bottom Row: K-factors for the same three points and for the same choice of cuts.
}\label{fig:dist_cthW}
\end{figure}

\begin{figure}
\centering
\includegraphics[width=0.49\linewidth]{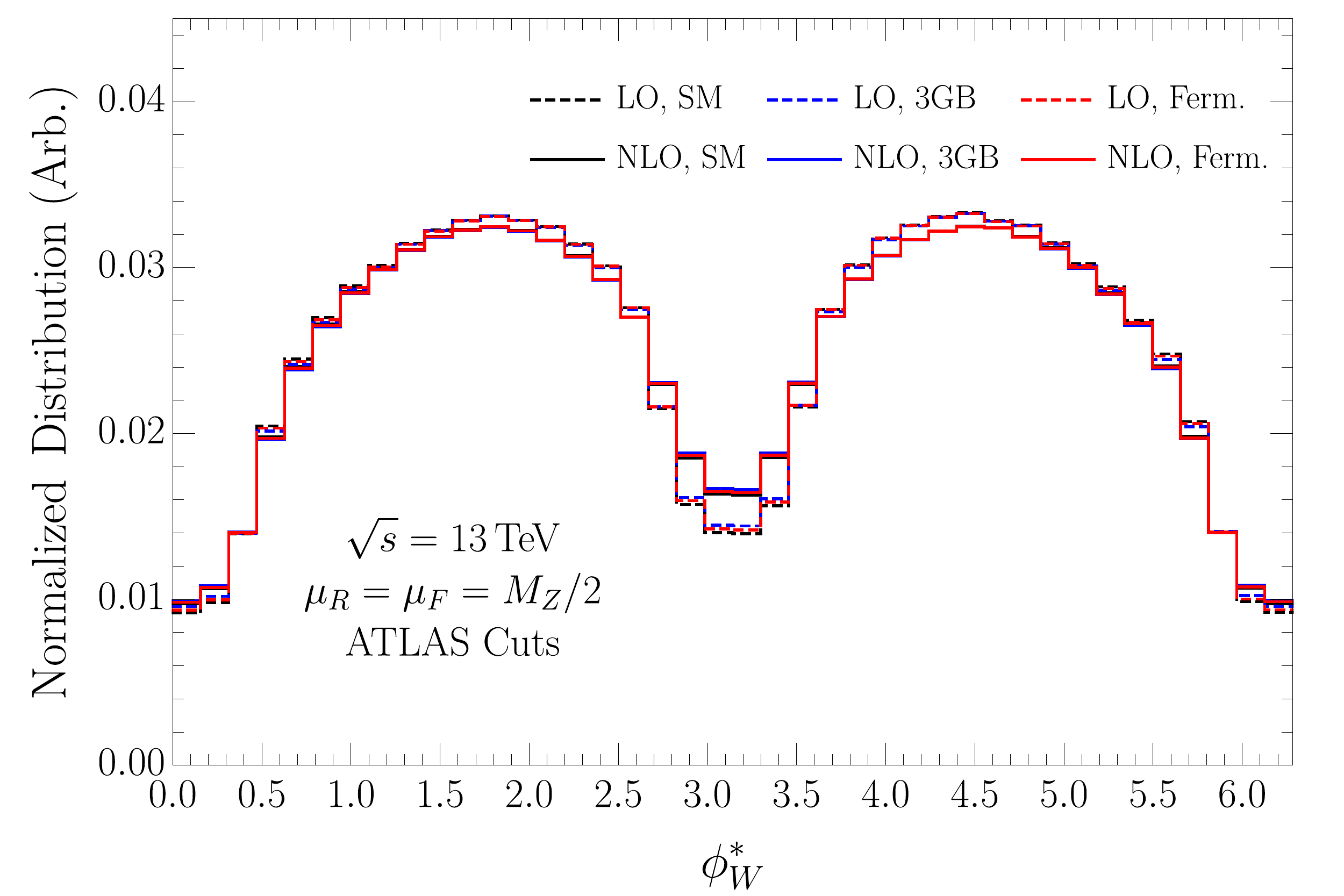}
\includegraphics[width=0.49\linewidth]{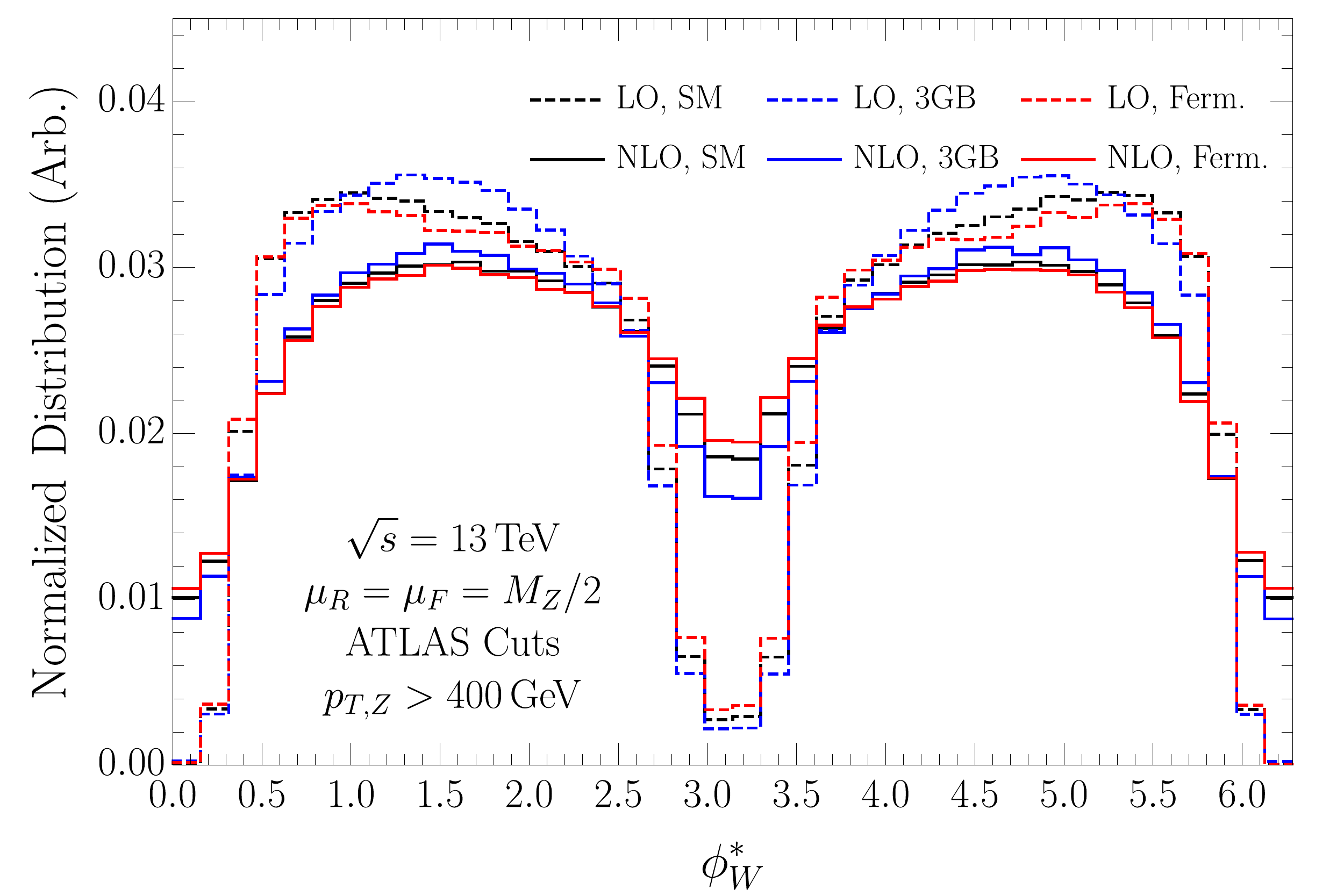}\\
\vspace{0.2cm}
\includegraphics[width=0.49\linewidth]{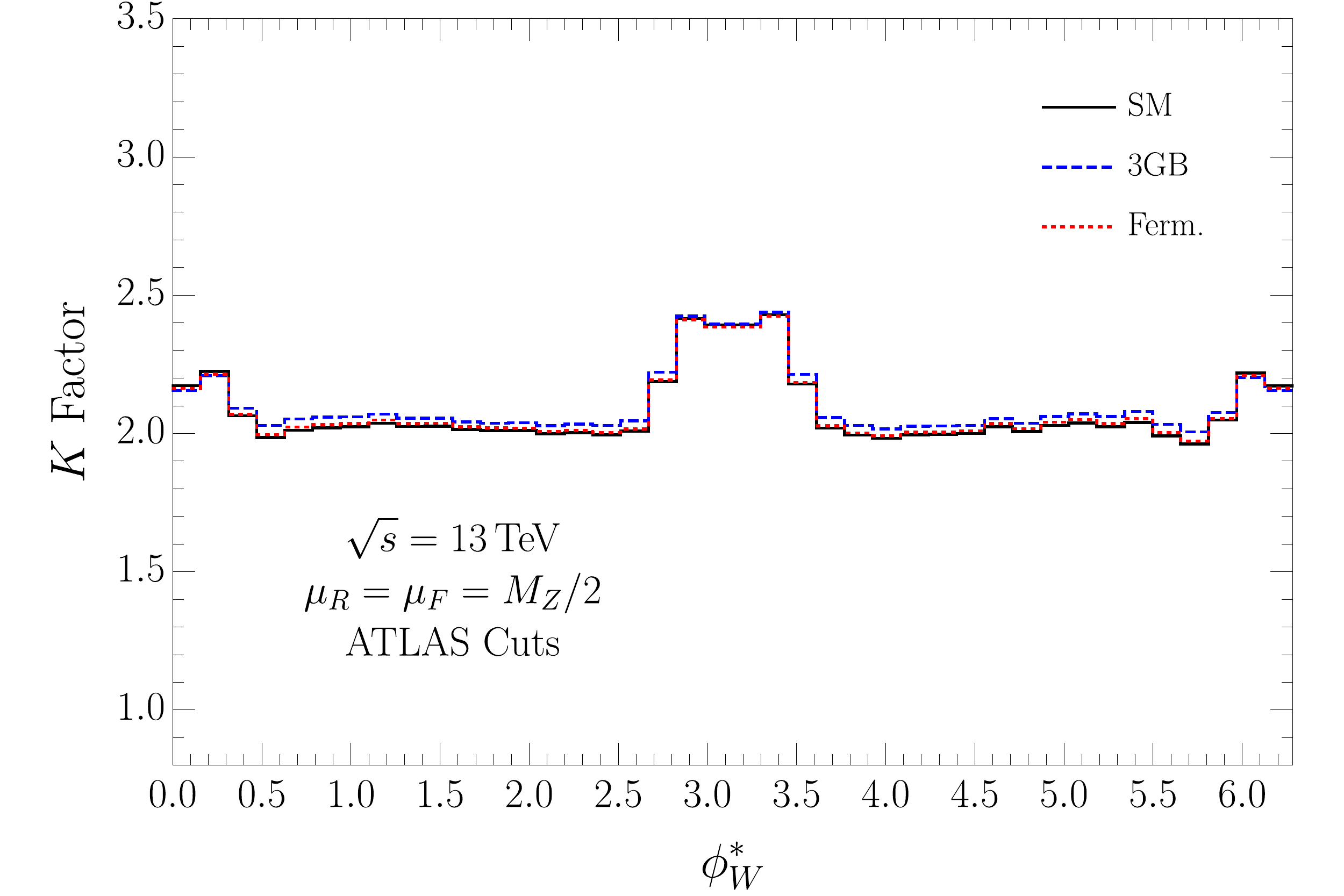}
\includegraphics[width=0.49\linewidth]{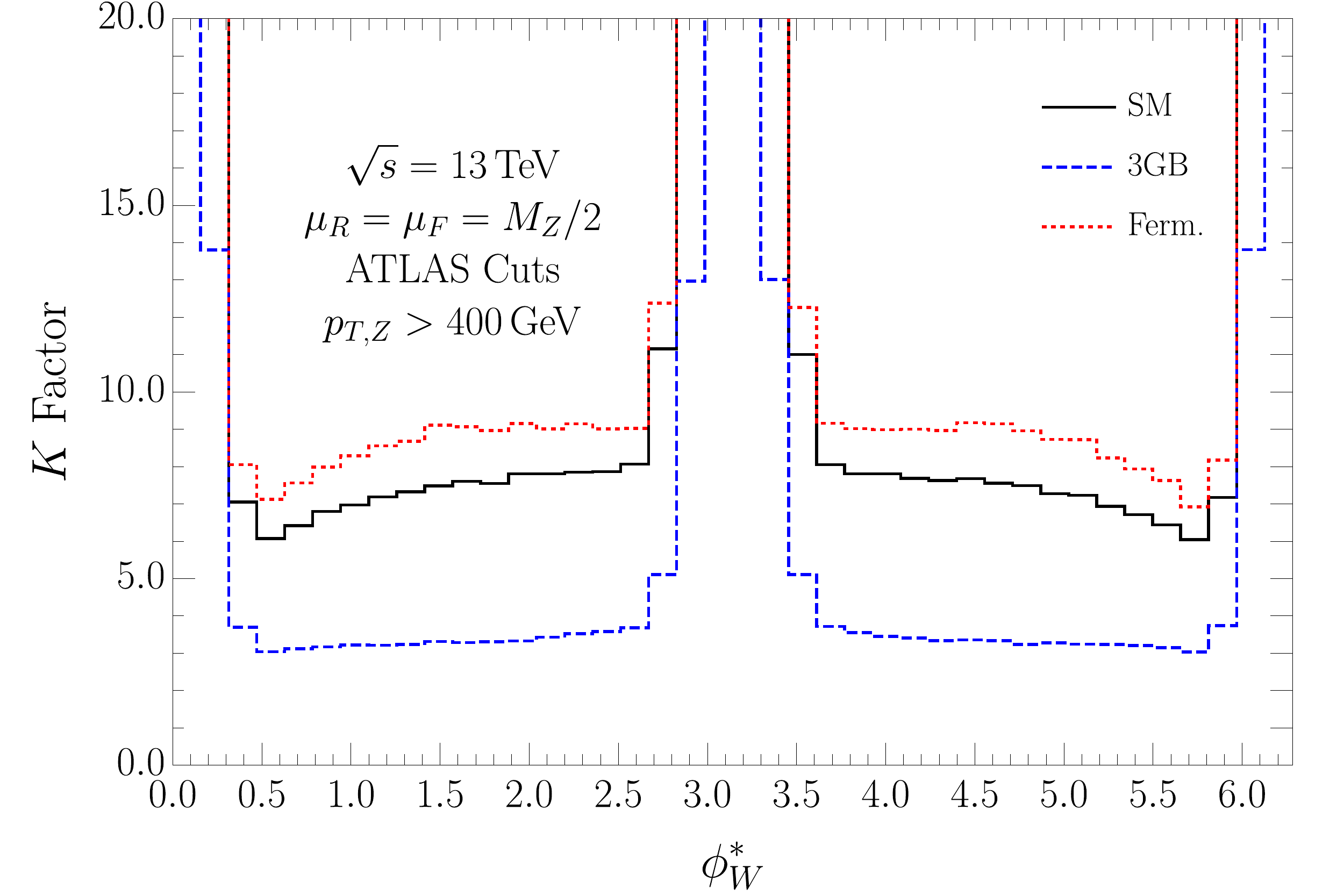}
\caption{
The same as Fig.~\ref{fig:dist_cthW}, but for the angular variable $\phi_W^*$.
}\label{fig:dist_phiW}
\end{figure}

We next consider the NLO effects on distributions of the angular
variables $\cos\theta^*_W$ and $\phi^*_W$. They are the angular
variables of the decayed charged lepton in the $W$ rest frame. We use
the helicity coordinate system as defined by
ATLAS~\cite{Aaboud:2019gxl}, in which the $z$ direction of the $W$ rest
frame is the $W$ direction-of-flight as seen in the $WZ$ center-of-mass
frame. The definitions of the $x$ and $y$ axes are given in
Ref.~\cite{Bern:2011ie} and a graphical representation is given in
Ref.~\cite{Baglio:2018rcu} (with a slight modification for the $z$
direction).
Angular variables in the decay products (particularly $\cos\theta_W^*$) are  useful for extracting maximal sensitivity to the gauge boson polarizations~\cite{Panico:2017frx}. 
As emphasized in Refs.~\cite{Falkowski:2016cxu, Panico:2017frx, Franceschini:2017xkh}, 
SMEFT effects lead to quadratic energy growth at the interference level only in the amplitude for producing two longitudinally polarized gauge bosons, Eq. \ref{eq:wzlims}. 
This behavior was exploited in Ref.~\cite{Franceschini:2017xkh} to maximize the sensitivity of the $p_{T,V}$ distribution to anomalous couplings.

In Figs.~\ref{fig:dist_cthW} and \ref{fig:dist_phiW} we show the normalized distributions of $\cos\theta_W^*$ and $\phi_W^*$ for the SM and
for  our two benchmark points at the  fiducial level (left) and with an additional cut requiring $p_{T,Z} > 400\,\mathrm{GeV}$ (right) to enhance the sensitivity of the distributions to the anomalous couplings.
Without the additional $p_{T,Z}$ cut, the distributions are quite insensitive to the small values of the anomalous couplings in our benchmark points. An  interesting effect of the NLO corrections is the washing out of the radiation zero present at  LO at $\cos\theta_W^* = -1$, evidenced by the large $K$~factor in this part of phase space.

After including the $p_{T,Z}$ cut, the LO samples are enriched with longitudinally polarized gauge bosons, and the distributions become much more sensitive to the anomalous couplings.
At NLO however, a great deal of this dependence is washed out as a result of the high $p_{T}$ bins  being more densely populated due to the real emission present at this order~\cite{Azatov:2017kzw, Franceschini:2017xkh}. While Ref.~\cite{Franceschini:2017xkh} suggested this could be ameliorated with a jet veto, it was also demonstrated in Ref.~\cite{Azatov:2017kzw} that a hard jet in the process is required to maintain access to the interference terms which grow quadratically with energy and are most sensitive to the SMEFT effects.   

\section{Fits}\label{sec:fits}

The results of Section~\ref{sec:fp} demonstrate that including  higher-order QCD effects in $WZ$ production in the presence of anomalous gauge and fermion couplings can lead to significantly different predictions than using the LO SMEFT calculation with the Standard Model $K$-factor.
We now consider how these effects change the observed limits on the anomalous couplings based on a fit to experimental data.
We consider the results in the case of a fit to only $WZ$ data, and then, as a step towards a global analysis, fit both $W^+W^-$ and $WZ$ data.

The existing experimental results on $W^+W^-$
 and $WZ$ production at both $8$ and $13\,\mathrm{TeV}$ are summarized in Table~\ref{tab:data}.
The $W^+W^-$ data from ATLAS collected at $8\,\mathrm{TeV}$ in Ref.~\cite{Aad:2016wpd} is systematically lower than the SM prediction in the lower bins, particularly in the 250 -- 350 GeV bin. We thus use only the highest bin in $p_T^{\mathrm{lead}}$ for our analysis.
The $W^+W^-$ data from CMS at $8\,\mathrm{TeV}$ in Ref.~\cite{Khachatryan:2015sga} includes both same and different flavor final states so there is a contribution from $ZZ$ production that we have not computed, so we do not include this result.
The ATLAS $13\,\mathrm{TeV}$ result with $3.16\,\mathrm{fb}^{-1}$ in Ref.~\cite{Aaboud:2017qkn} uses data that is also included in the updated result with $36.1\,\mathrm{fb}^{-1}$~\cite{Aaboud:2019nkz}, so we will exclude this result as well.

To perform the fits, we construct a $\chi^2$ function with the data from the remaining six data sets from Refs.~\cite{Aaboud:2019nkz, Aad:2016ett, Khachatryan:2016poo, Aaboud:2019gxl, Sirunyan:2019bez}, using the distributions indicated in Table~\ref{tab:data}.
The data in the distributions for Refs.~\cite{Aaboud:2019nkz, Aad:2016ett, Khachatryan:2016poo, Aaboud:2019gxl}, including both statistical and systematic uncertainties was obtained from the corresponding supplementary information.
 We combine the different sources of uncertainty in quadrature in each bin, neglecting any correlations.
The data in Refs.~\cite{Aad:2016wpd, Sirunyan:2019bez} is not available online, so we digitize the plots to obtain the observed data and statistical uncertainties and add an additional $5\%$ systematic uncertainty bin-by-bin, again neglecting correlations.
In each case, the Standard Model prediction for the $W^+W^-$ or $WZ$ contribution for each distribution was found by digitizing the plots in the experimental papers.
To account for detector effects, we normalize our theory predictions bin-by-bin to agree with the Standard Model predictions taken from the experimental results.

\begin{table}
\begin{center}
\begin{tabular}{ l|lcll}
\hline
Channel  & Distribution & \# bins   &\hspace*{0.2cm} Data set & \hspace*{0.2cm}Int.  Lum.
\\ 
\hline
$WW\rightarrow \ell^+\ell^{\prime -}+\sla{E}_T\; (0j)$ & $p^{\rm leading, lepton}_{T}$, Fig.~11 & 1 
& ATLAS 8 TeV &20.3 fb$^{-1}$~\cite{Aad:2016wpd}
\\[0mm]
$WW\rightarrow e^\pm \mu^\mp+\sla{E}_T\; (0j)$ &  $p_T^{leading ,lepton}$, Fig.~7 & 5 
& ATLAS 13 TeV &36.1 fb$^{-1}$~\cite{Aaboud:2019nkz}
\\[0mm]
$WZ\rightarrow \ell^+\ell^{-}\ell^{(\prime)\pm}$ & $m_{T}^{WZ}$, Fig.~5 & 2 
& ATLAS 8 TeV & 20.3 fb$^{-1}$~\cite{Aad:2016ett}
\\[0mm]
$WZ\rightarrow \ell^+\ell^{-}\ell^{(\prime)\pm}+\sla{E}_T$ 
& $Z$ candidate $p_{T}^{\ell\ell}$, Fig.~5 & 9 
& CMS 8 TeV &19.6 fb$^{-1}$~\cite{Khachatryan:2016poo}
\\[0mm]
$WZ\rightarrow \ell^+\ell^{-}\ell^{(\prime)\pm}$ &  $m_{T}^{WZ}$ Fig.~4c & 6
& ATLAS 13 TeV &36.1 fb$^{-1}$~\cite{Aaboud:2019gxl}
\\[0mm]
$WZ\rightarrow \ell^+\ell^{-}\ell^{(\prime)\pm}+\sla{E}_T$ & $m^{WZ}$, Fig.~15a & 3
& CMS 13 TeV, &35.9 fb$^{-1}$~\cite{Sirunyan:2019bez}
\\[0mm]
\hline
\end{tabular}
\caption{
Experimental data included in our study.
The third column shows the number of bins used in our analysis, always counting from the highest.
}\label{tab:data}
\end{center}
\end{table}

\subsection{Fits to $WZ$ Data}\label{ss:wz_fits}

\begin{figure}
\centering
\subfigure{\includegraphics[width=0.4\linewidth]{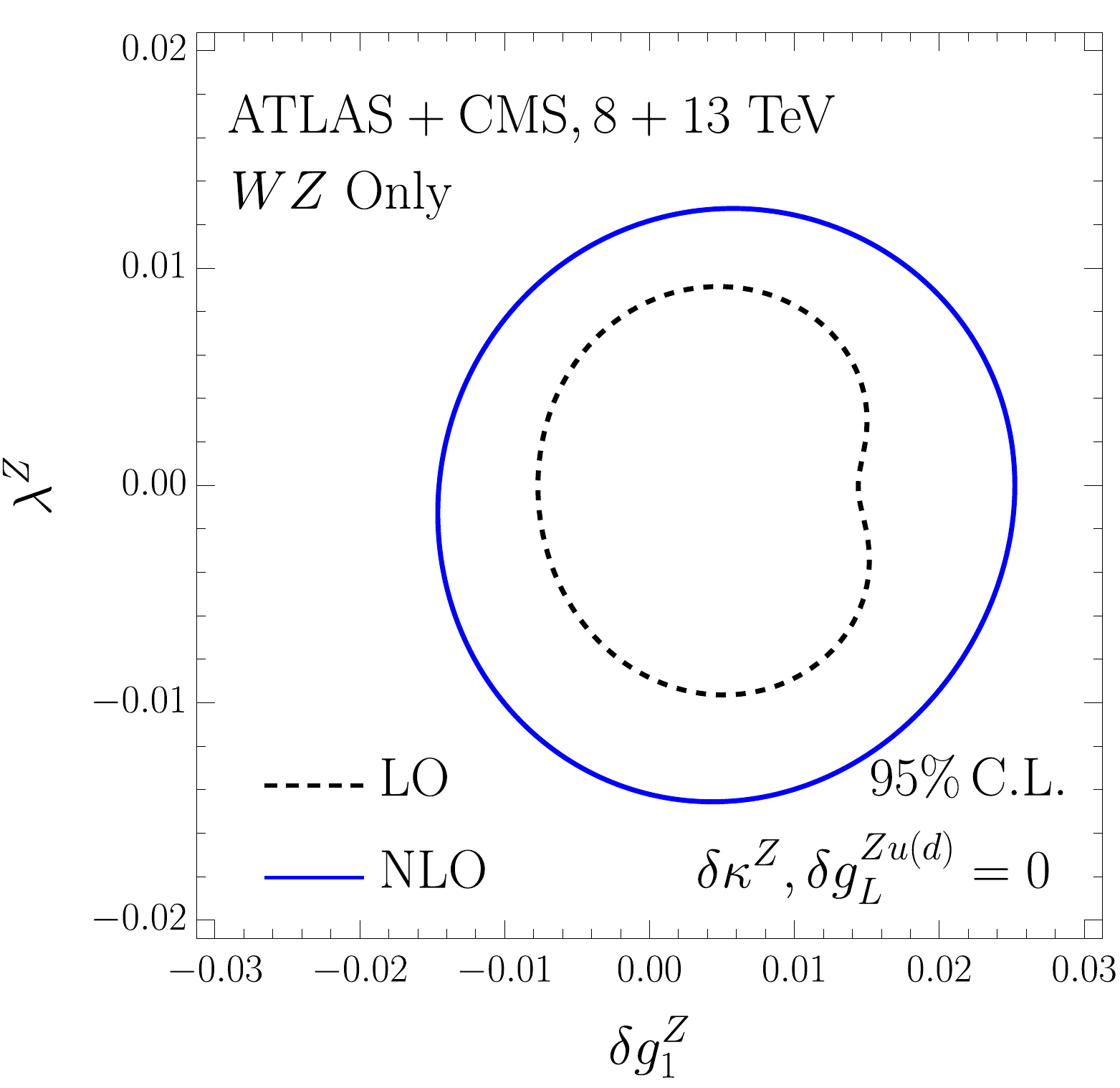}}~~~
\subfigure{\includegraphics[width=0.4\linewidth]{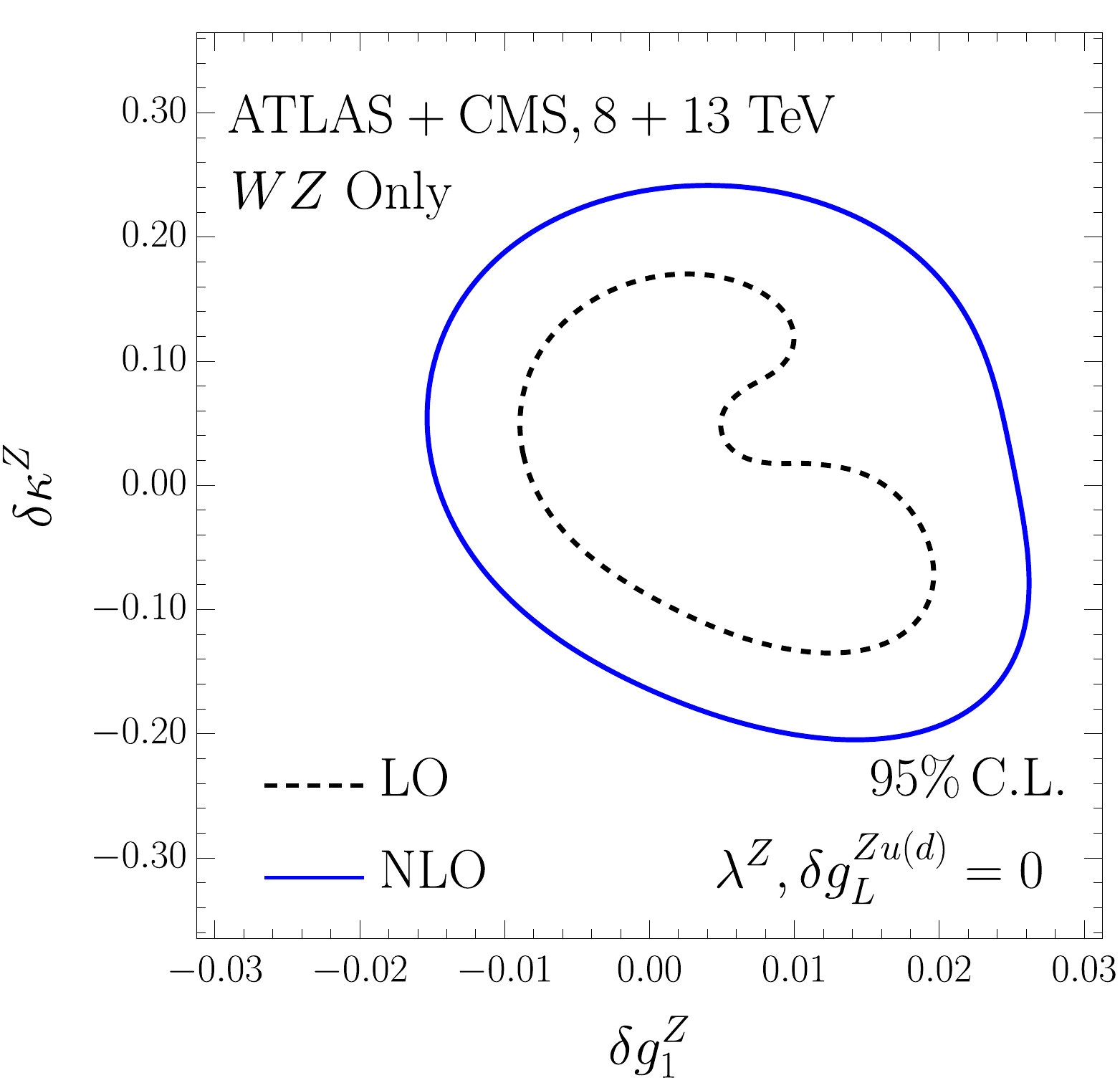}}\qquad~~~\\
\subfigure{\includegraphics[width=0.4\linewidth]{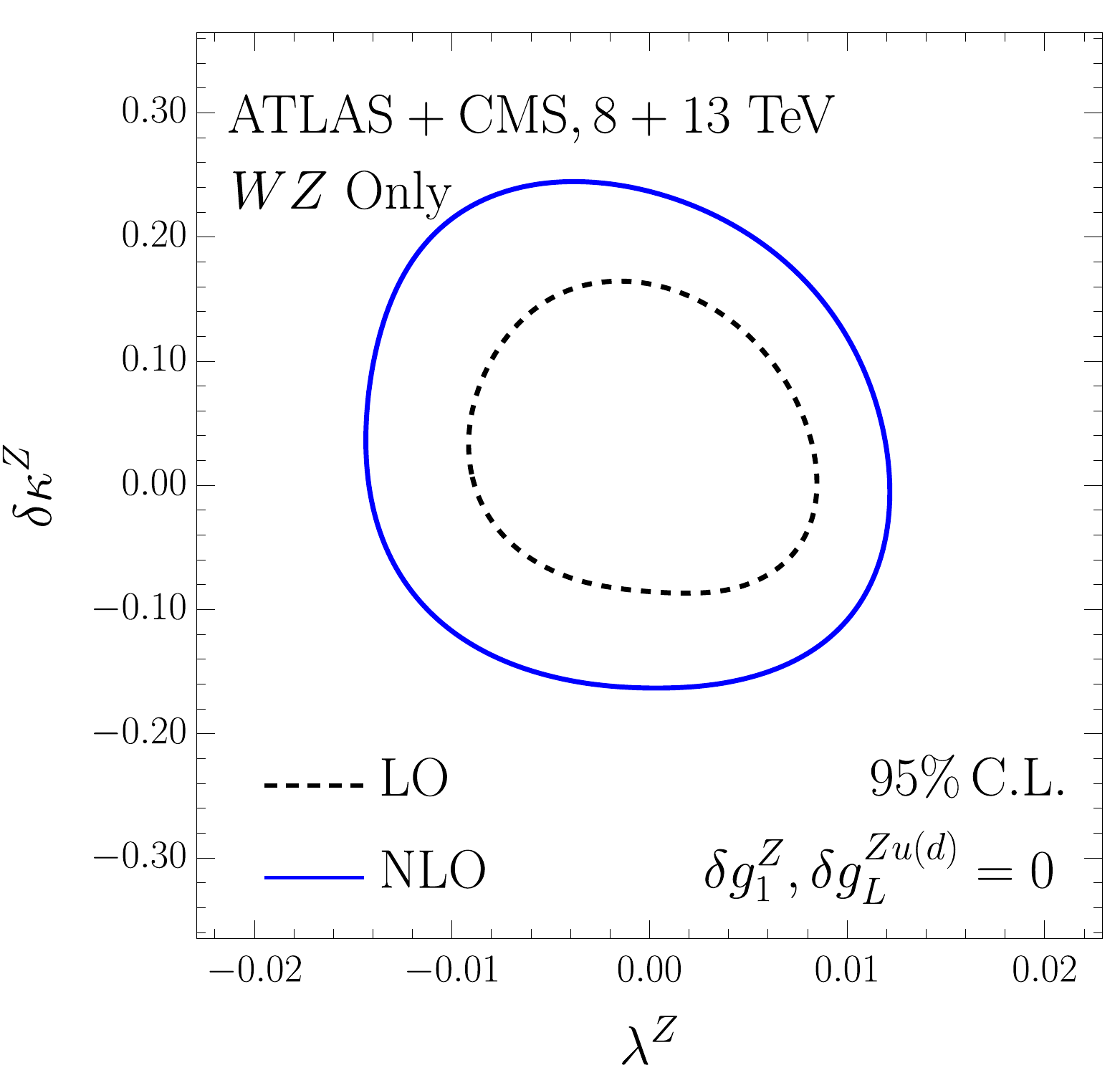}}~~~
\subfigure{\includegraphics[width=0.4\linewidth]{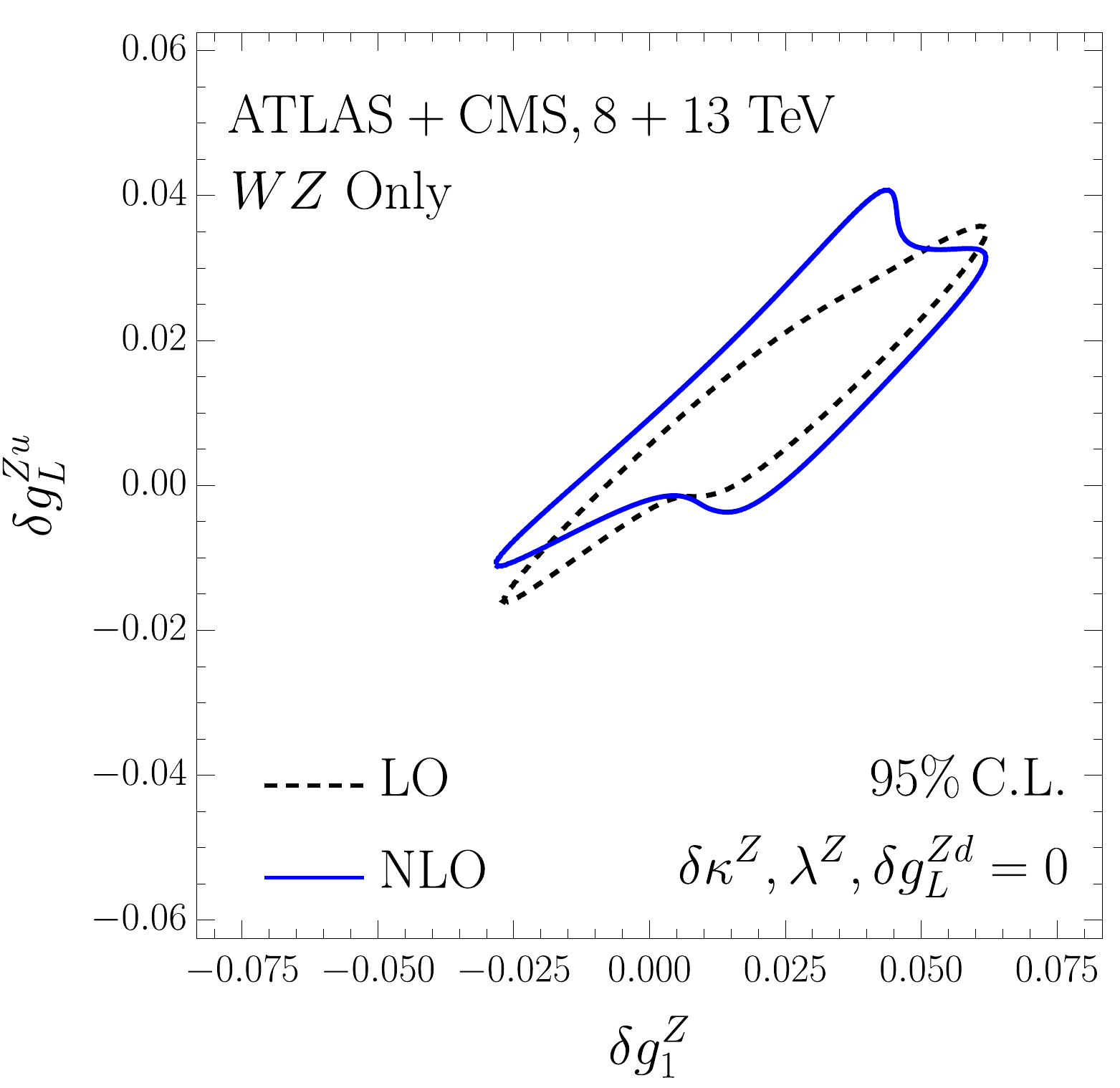}}\qquad~~~\\
\subfigure{\includegraphics[width=0.4\linewidth]{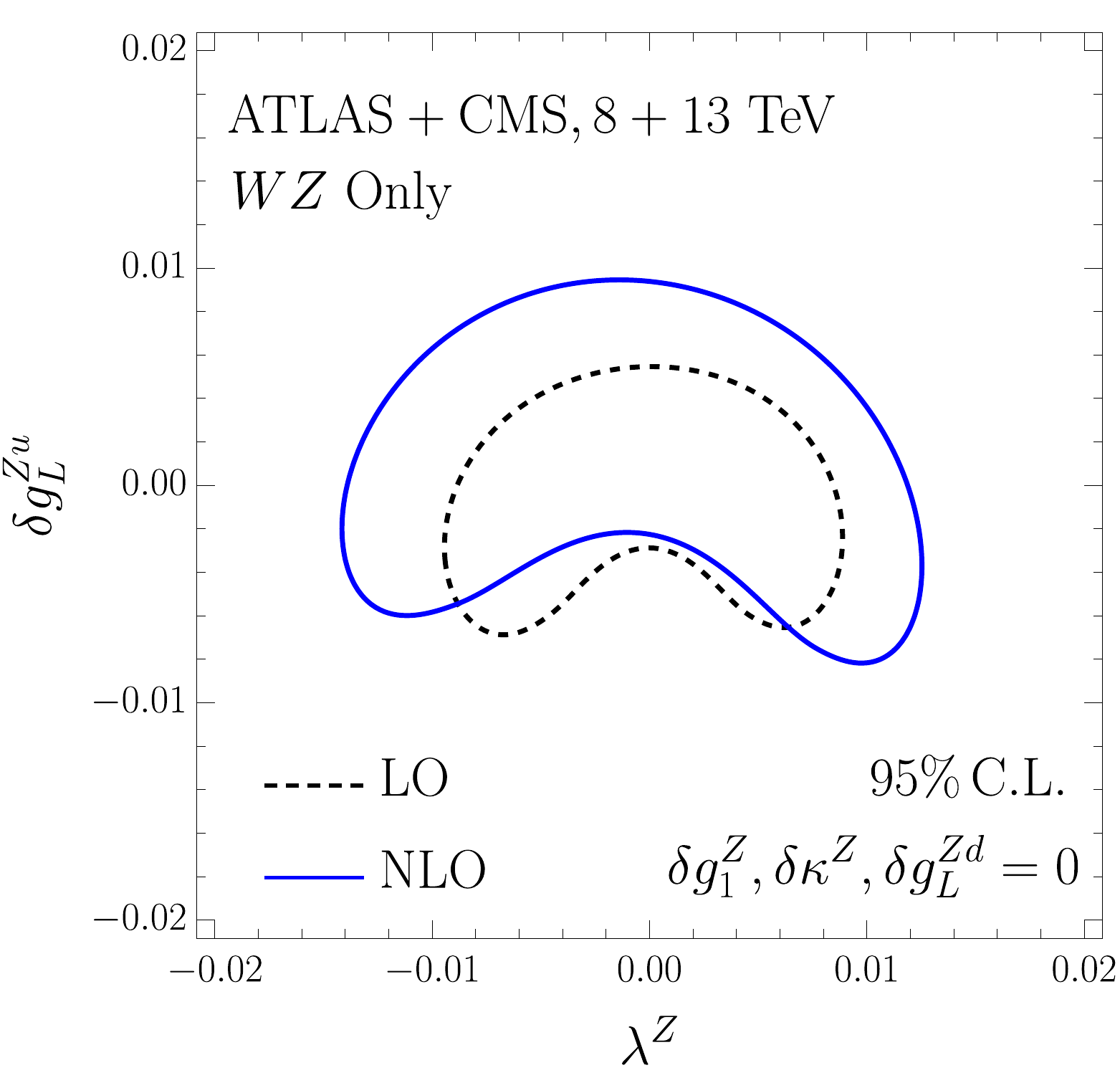}}~~~
\subfigure{\includegraphics[width=0.4\linewidth]{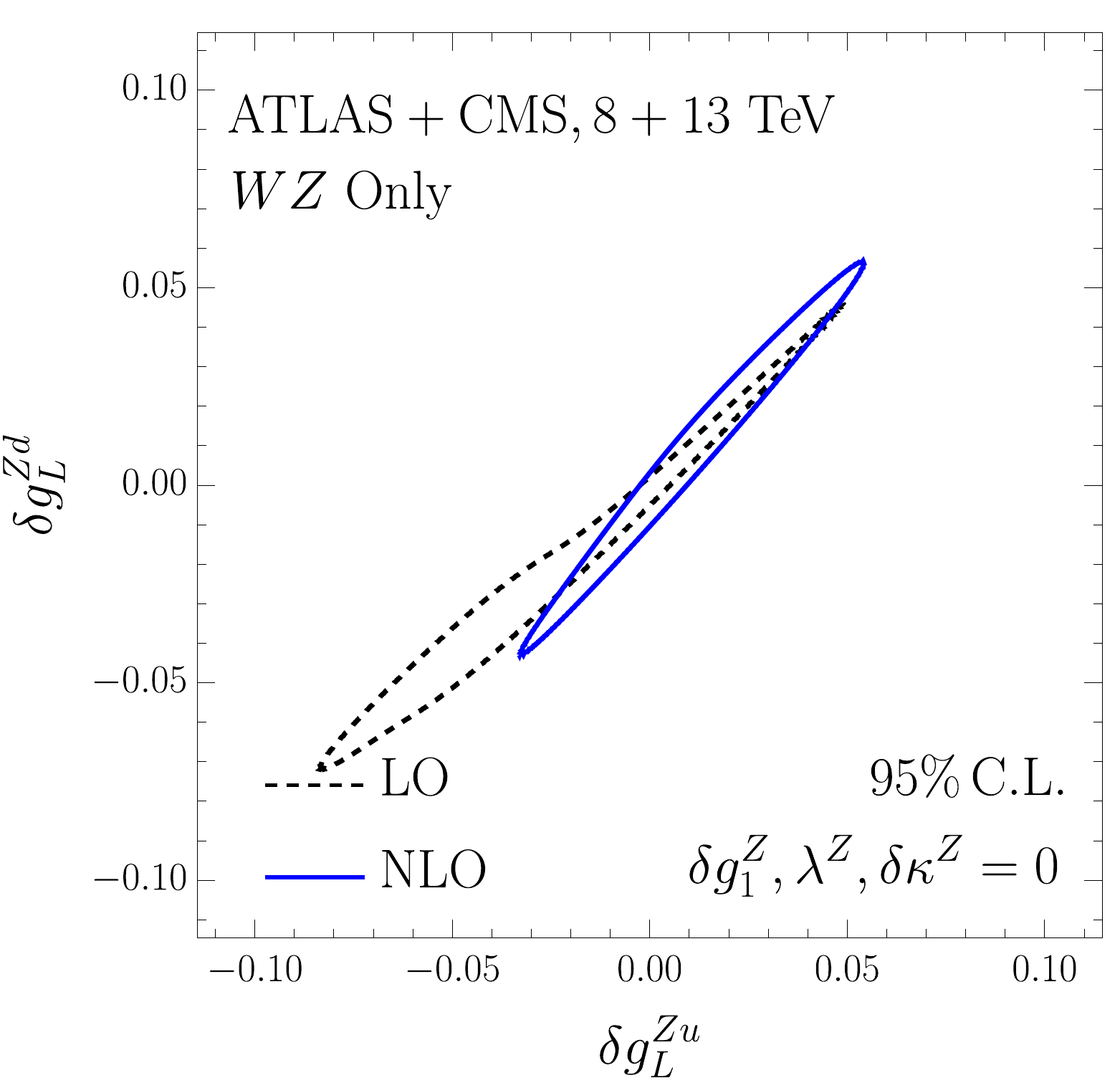}}\qquad~~~
\vskip -0.5cm
\caption{
95\% C.L. allowed regions for different combinations of anomalous gauge and fermion couplings based on a fit to the $8$ and $13\,\mathrm{TeV}$ $WZ$ data from ATLAS and CMS~\cite{Aad:2016ett, Khachatryan:2016poo, Aaboud:2019gxl, Sirunyan:2019bez}.
The results with the SMEFT treated at leading order (LO) are shown as dashed black contours and the constraints with the SMEFT treated at next-to-leading order  (NLO) are shown in solid blue. 
In each panel we set the three couplings not shown to zero.
}\label{fig:wz_proj_contours}
\end{figure}

We first present the fits to the 8 and $13\,\mathrm{TeV}$ $WZ$ data from ATLAS and CMS~\cite{Aad:2016ett, Khachatryan:2016poo, Aaboud:2019gxl, Sirunyan:2019bez}. In Fig.~\ref{fig:wz_proj_contours} we show the 95\% C.L. allowed regions from various two parameter fits to the anomalous couplings, in each case fixing the other three couplings to zero. 
As anticipated in Section~\ref{sec:fp}, the constraints using the LO and NLO predictions for the SMEFT contributions are quite different. The constraints on the different combinations of gauge couplings are weaker, in some directions by a factor of two. This is consistent with the behavior of the distributions with our ``Gauge" benchmark point in Fig.~\ref{fig:dists}. The effect is somewhat less dramatic in the case of anomalous fermion couplings, but there is still a large difference between the limits at LO and NLO.

In Section~\ref{sec:eftres}, we noted that the helicity amplitudes for $WZ$ production had only a sub-leading (in $s/M_Z^2$)  dependence on $\delta\kappa^Z$ in the high energy limit.
  Measurements of $WZ$ production are thus much less sensitive to $\delta\kappa^Z$, and we see in Fig.~\ref{fig:wz_proj_contours} that the limits on $\delta \kappa^Z$ are indeed an order of magnitude weaker than those on $\delta g_1^Z$ and $\lambda^Z$.
We also note that there is a near flat direction in the $\delta g_1^Z$ -- $\delta g_L^{Zu}$ plane, and an even more robust flat direction in the  $\delta g_L^{Zu}$ -- $\delta g_L^{Zd}$ plane, in agreement with the scalings in Eq.~\ref{eq:wzlims}.

\subsection{Combined Fits to $W^+W^-$ and $WZ$ Data}\label{ss:comb_fits}

In the previous section, it was demonstrated that treating the SMEFT consistently at NLO significantly changes the anomalous coupling constraints using $WZ$ data only. On the other hand, in Ref.~\cite{Baglio:2018bkm}, it was shown that the NLO effects on $W^+W^+$ distributions in the presence of anomalous couplings are relatively mild --- in other words, using the $K$-factor derived at the SM is an adequate approximation for setting limits.  It is of interest to understand to what extent the significant changes between LO and NLO fits in Fig.~\ref{fig:wz_proj_contours} remain when including $W^+W^-$ data.
Note that this is only a first step: the anomalous couplings are also constrained by other measurements both in Higgs data, top
quark physics and at LEP.

\begin{figure}[ht!]
\centering
\subfigure{\includegraphics[width=0.4\linewidth]{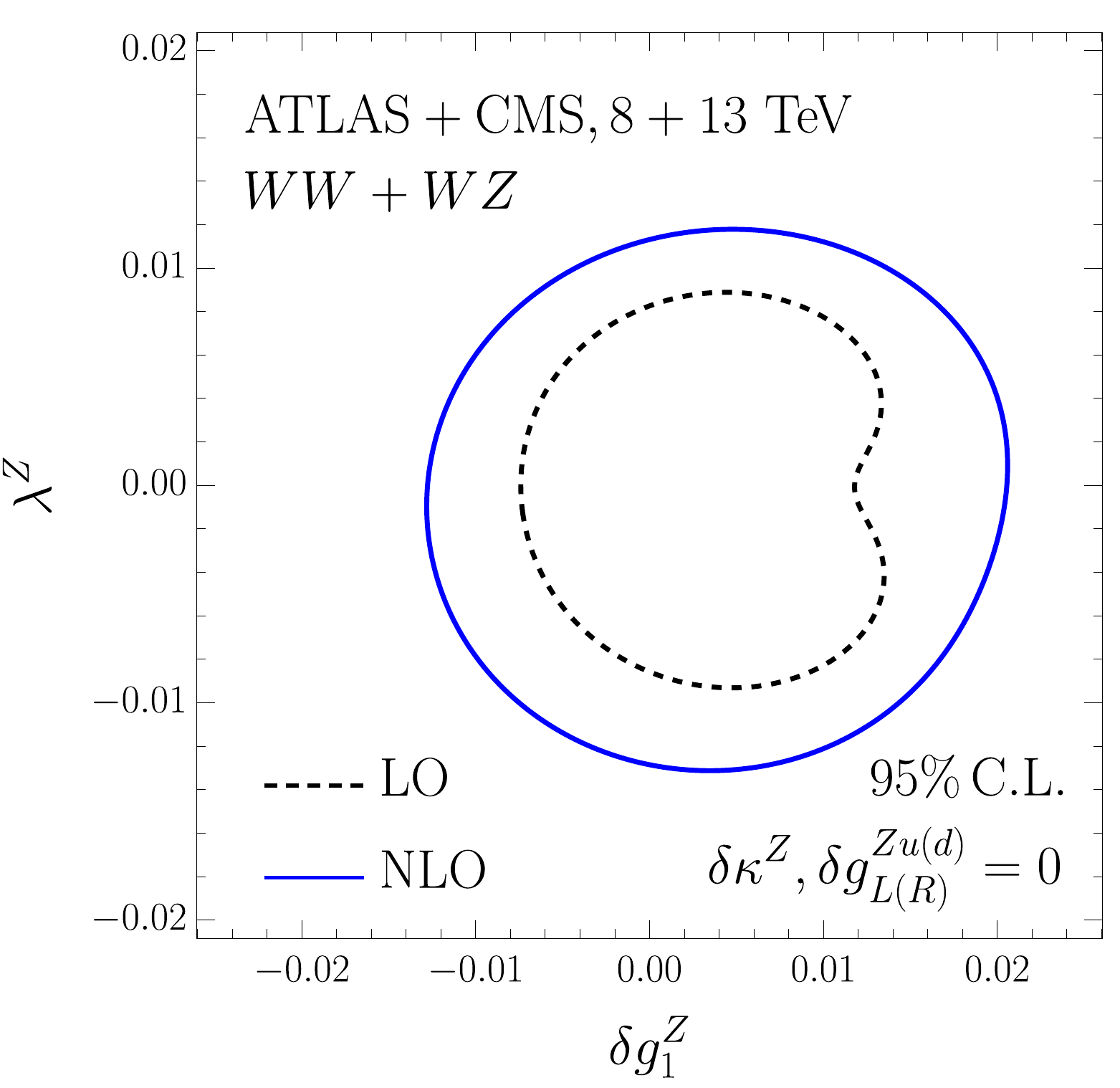}}~~~
\subfigure{\includegraphics[width=0.4\linewidth]{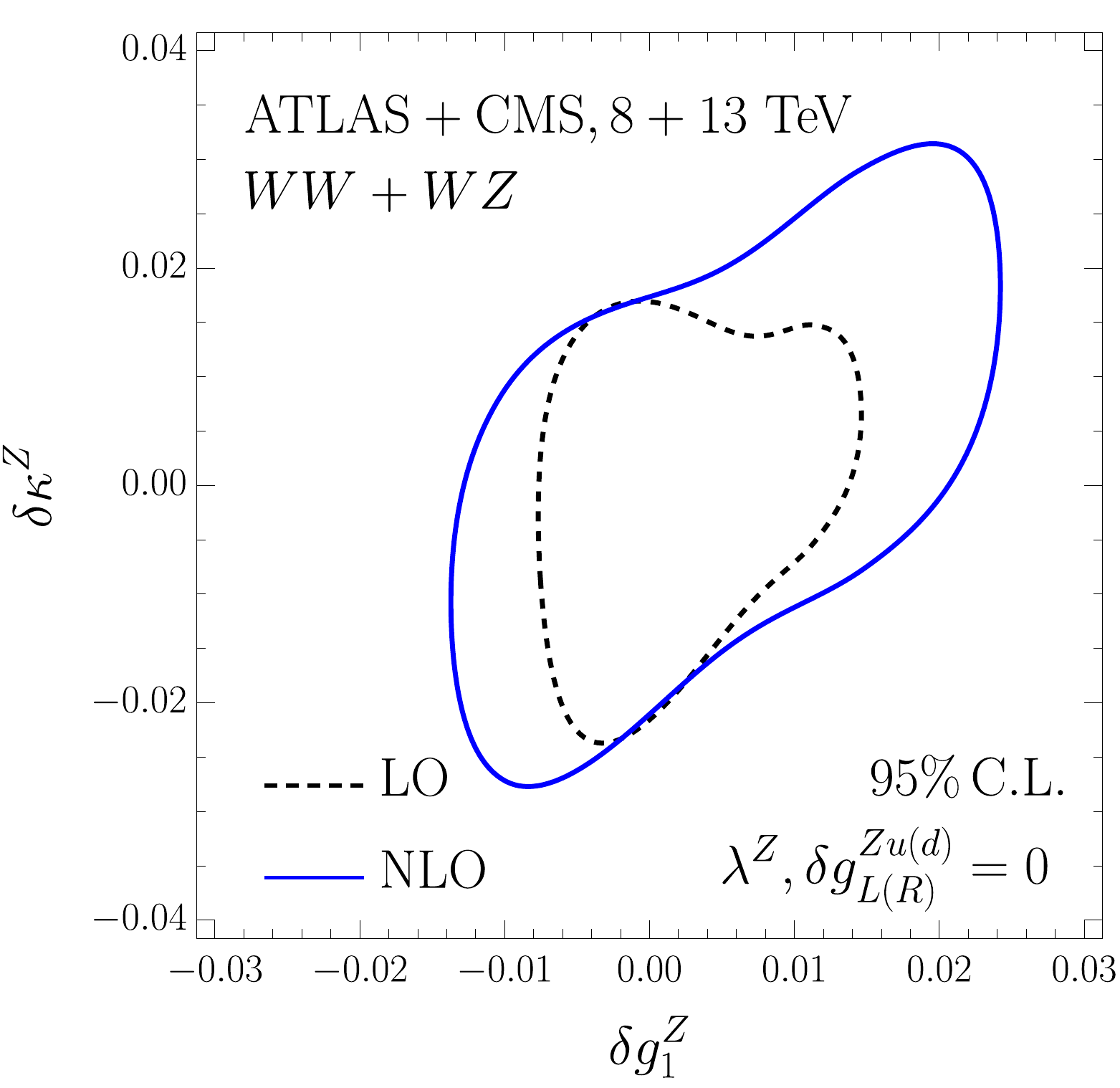}}\qquad~~~\\
\subfigure{\includegraphics[width=0.4\linewidth]{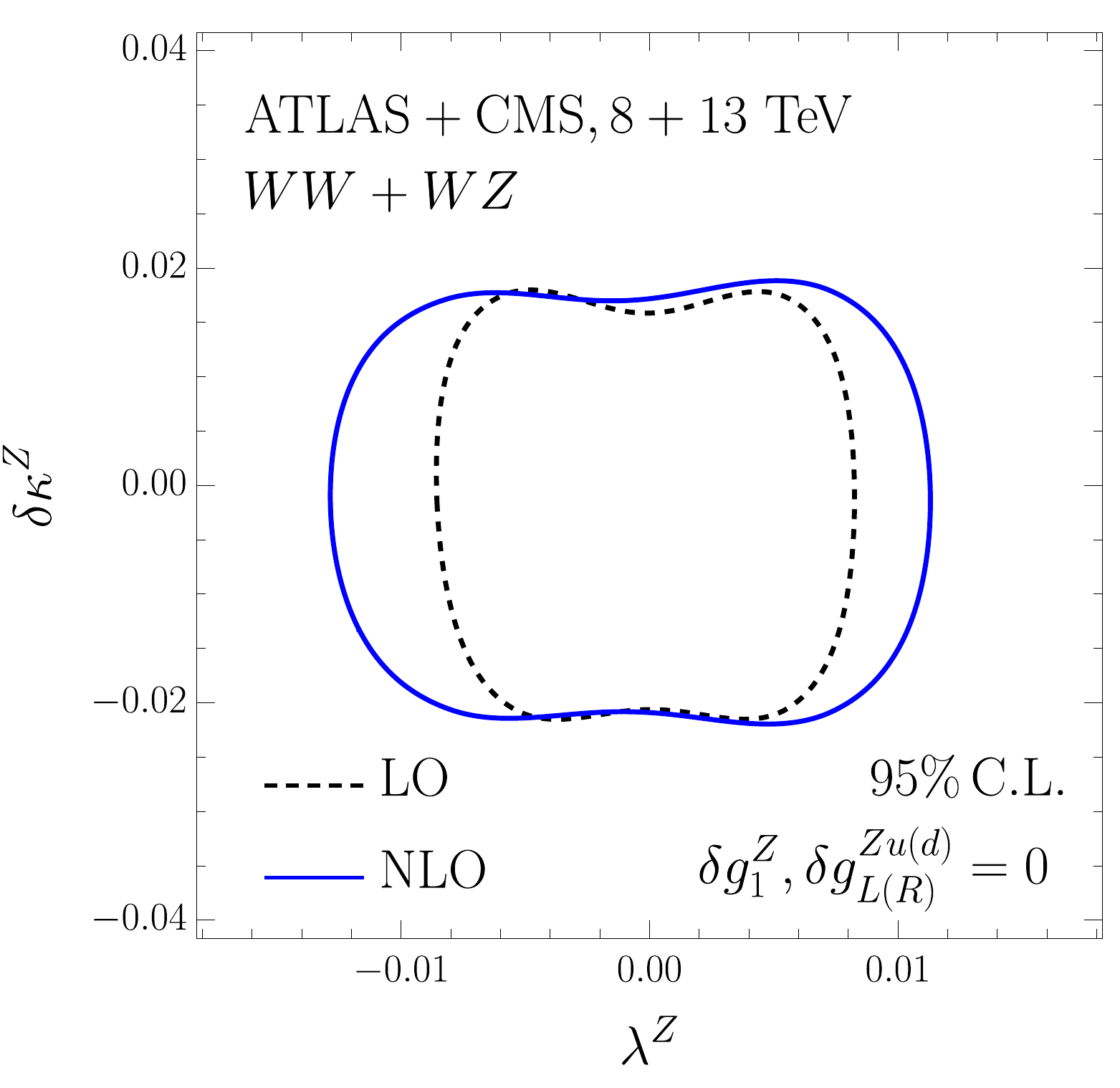}}~~~
\subfigure{\includegraphics[width=0.4\linewidth]{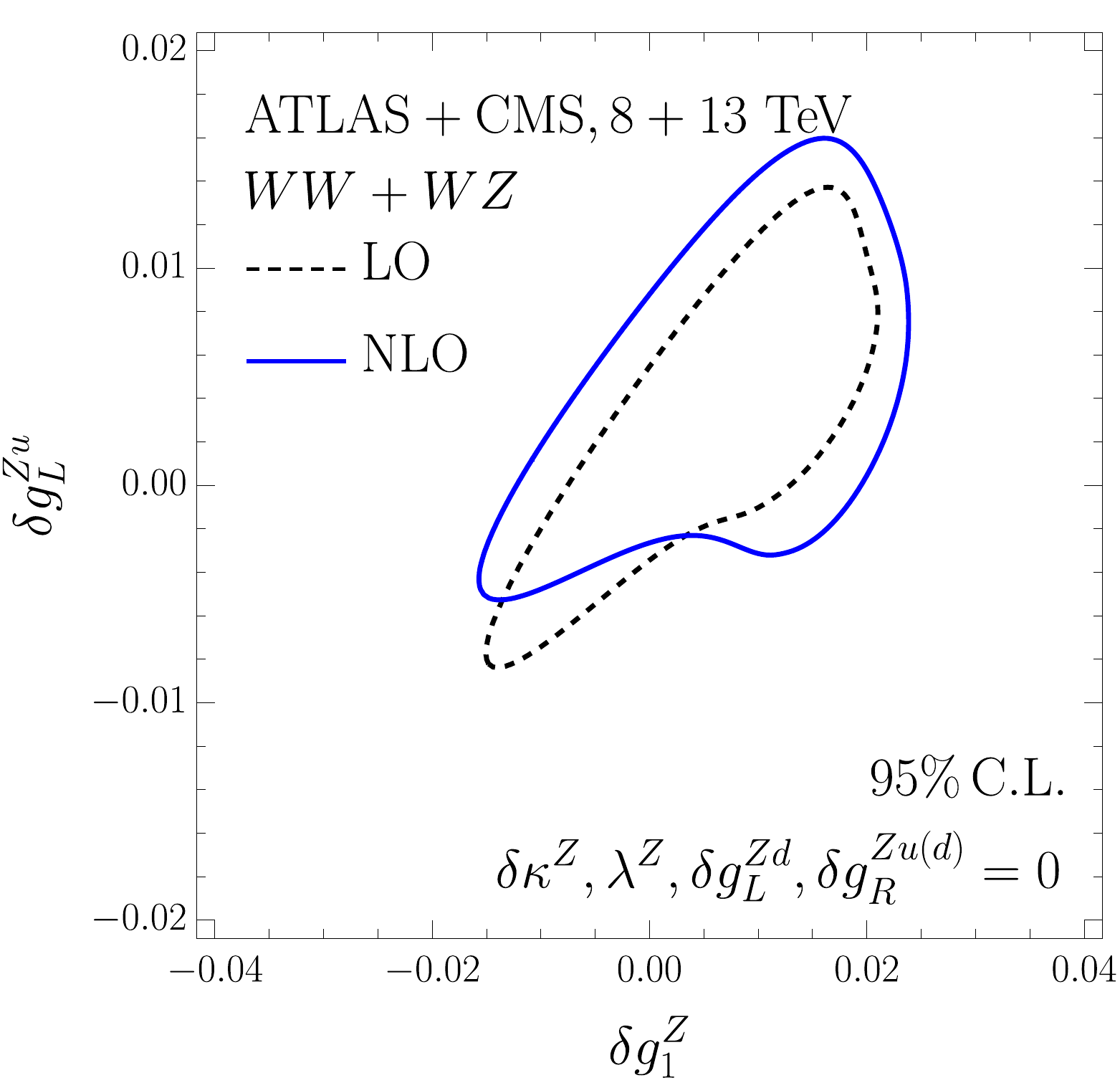}}\qquad~~~\\
\subfigure{\includegraphics[width=0.4\linewidth]{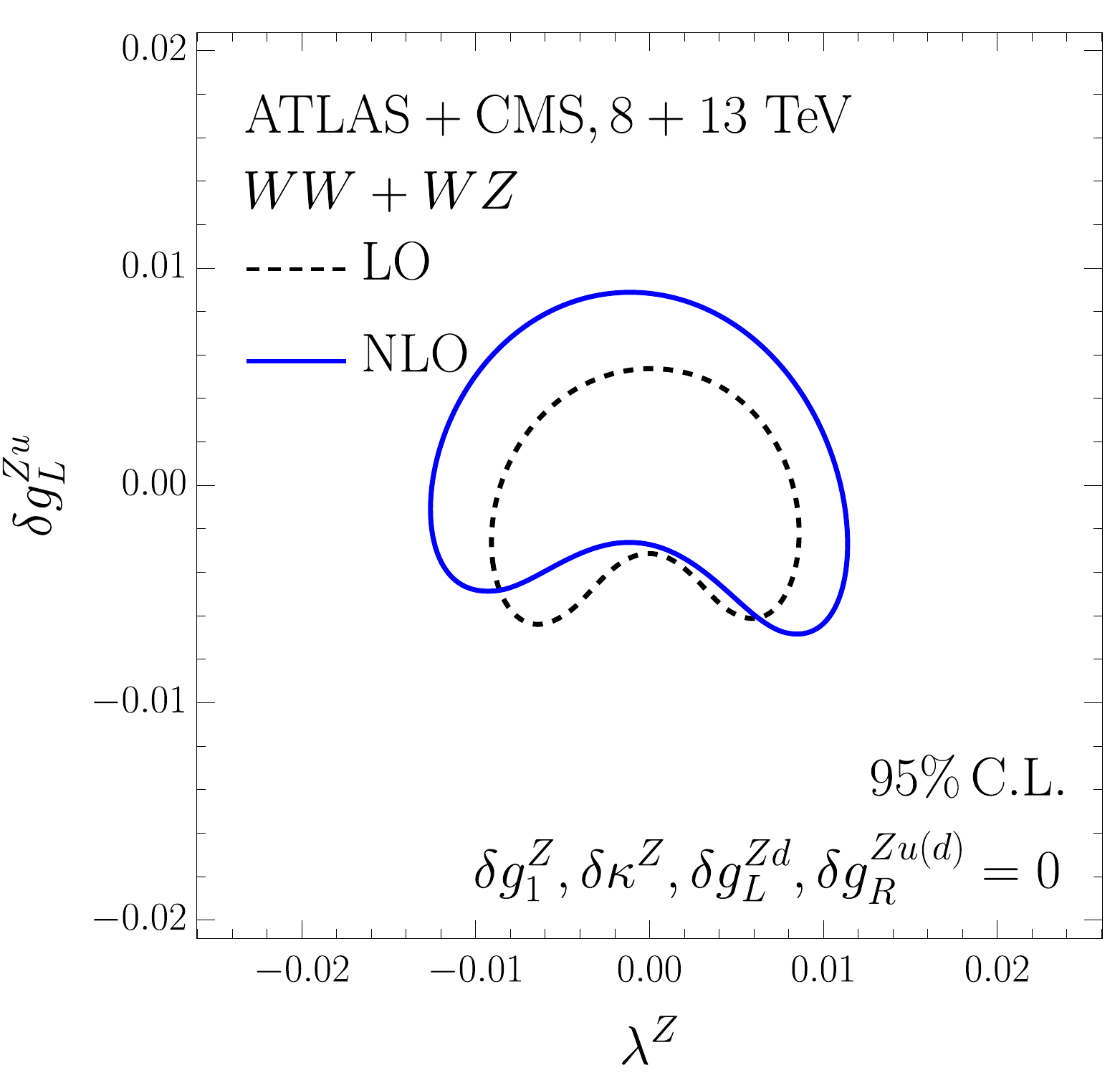}}~~~
\subfigure{\includegraphics[width=0.4\linewidth]{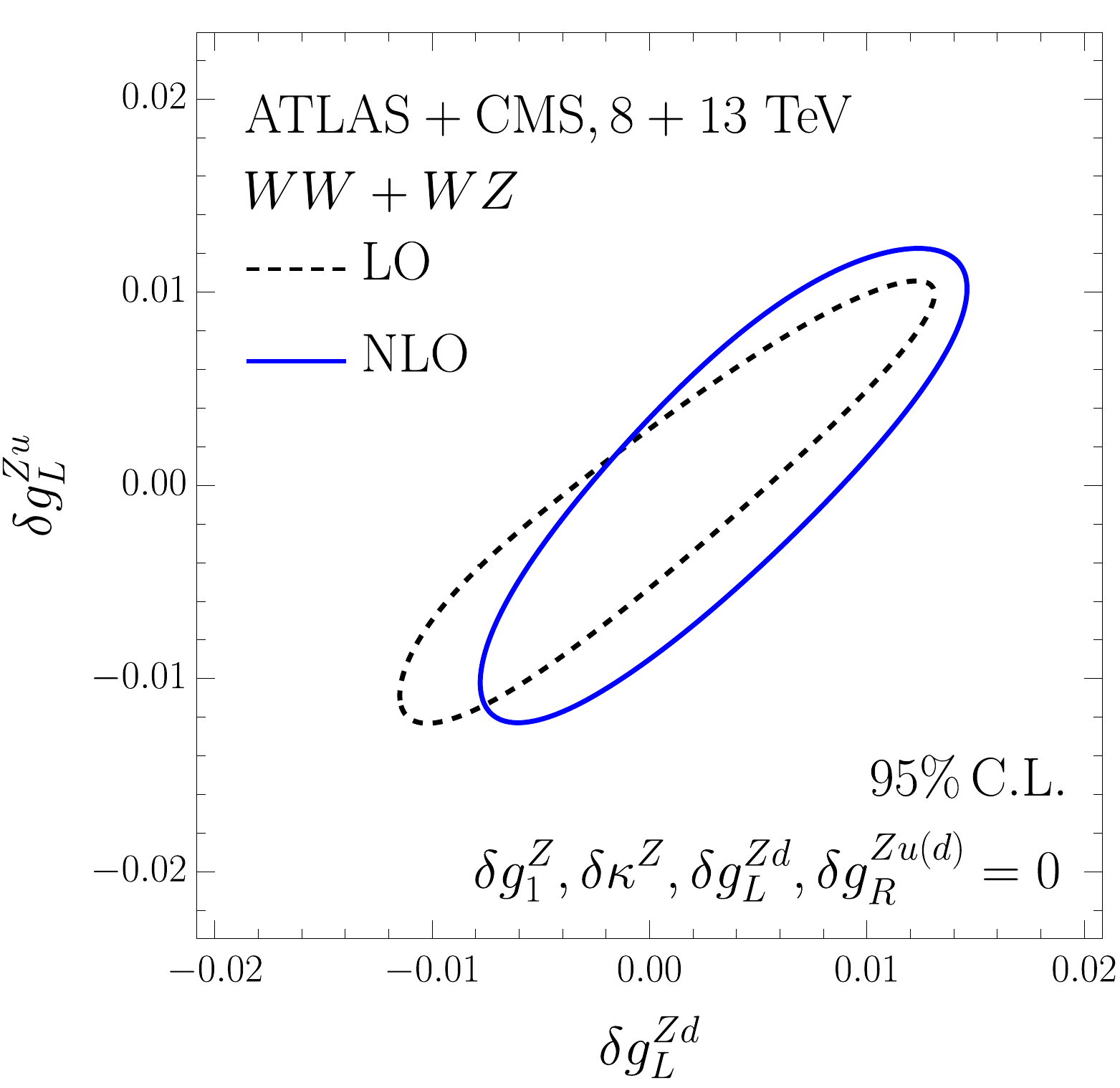}}\qquad~~~
\vskip -0.5cm
\caption{
As in Fig.~\ref{fig:wz_proj_contours}, but using both $WW$ and $WZ$ data.
}\label{fig:all_proj_contours}
\end{figure}

In Fig.~\ref{fig:all_proj_contours}, we consider the results for various combinations of couplings with the same setup as in Fig.~\ref{fig:wz_proj_contours}, with the other anomalous couplings fixed to zero. 
The most obvious result is that, even when combined with $W^+W^-$ data, the effects of treating the SMEFT at NLO in $WZ$ 
and $W^+W^-$ production on the limits are still quite substantial in many directions in parameter space.
The first panel is clearly mostly constrained by $WZ$ data.
As discussed in Subsection~\ref{ss:wz_fits}, $\delta\kappa^Z$ is much better constrained when $W^+W^-$ data is included, and since the NLO effects in $W^+W^-$ production are very small, the limits in the $\delta \kappa^Z-\lambda^Z$ plane (with all other couplings fixed to zero) are quite
similar at LO and NLO. In the $\delta g_1^Z$ -- $\delta\kappa^Z$ plane, however, there is a flat direction in $W^+W^-$ production in the high energy limit (Eq. \ref{eq:wwlims}), which is broken by the $WZ$ data and the NLO effects are significant.


We have computed the rates up to quadratic order, $\mathcal{O}(1/\Lambda^4)$. 
This  has a theoretical complication, however, because in principle
dimension-8 operators may contribute at the same order in
$1/\Lambda^4$ in the most general EFT framework and
these effects are not considered here, as we have
implicitely assumed that the contributions from the
$\mathcal{O}(1/\Lambda^4)$ dimension-8 operators in the Lagrangian are subleading.
We discuss the effects and inherent assumptions involved in truncating
the SMEFT expansion explicitly at $\mathcal{O}(1/\Lambda^2)$, i.e.,
treating the anomalous couplings only to linear order, in Appendix B.
This discussion shows that the $\mathcal{O}(1/\Lambda^4)$ terms in
the cross section are the leading effect in our fits.

\begin{figure}[h]
\centering
\subfigure{\includegraphics[width=0.4\linewidth]{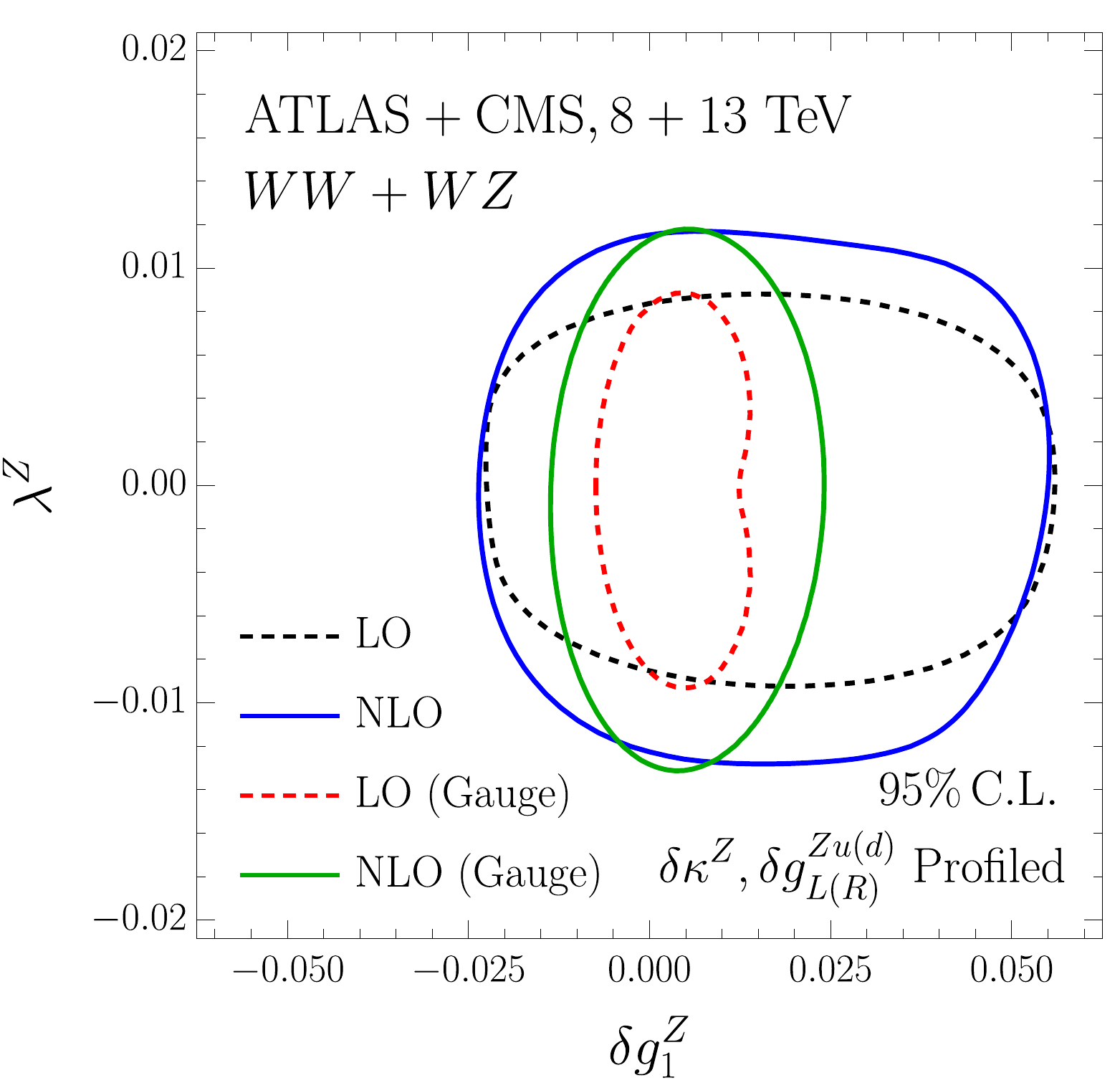}}~~~
\subfigure{\includegraphics[width=0.4\linewidth]{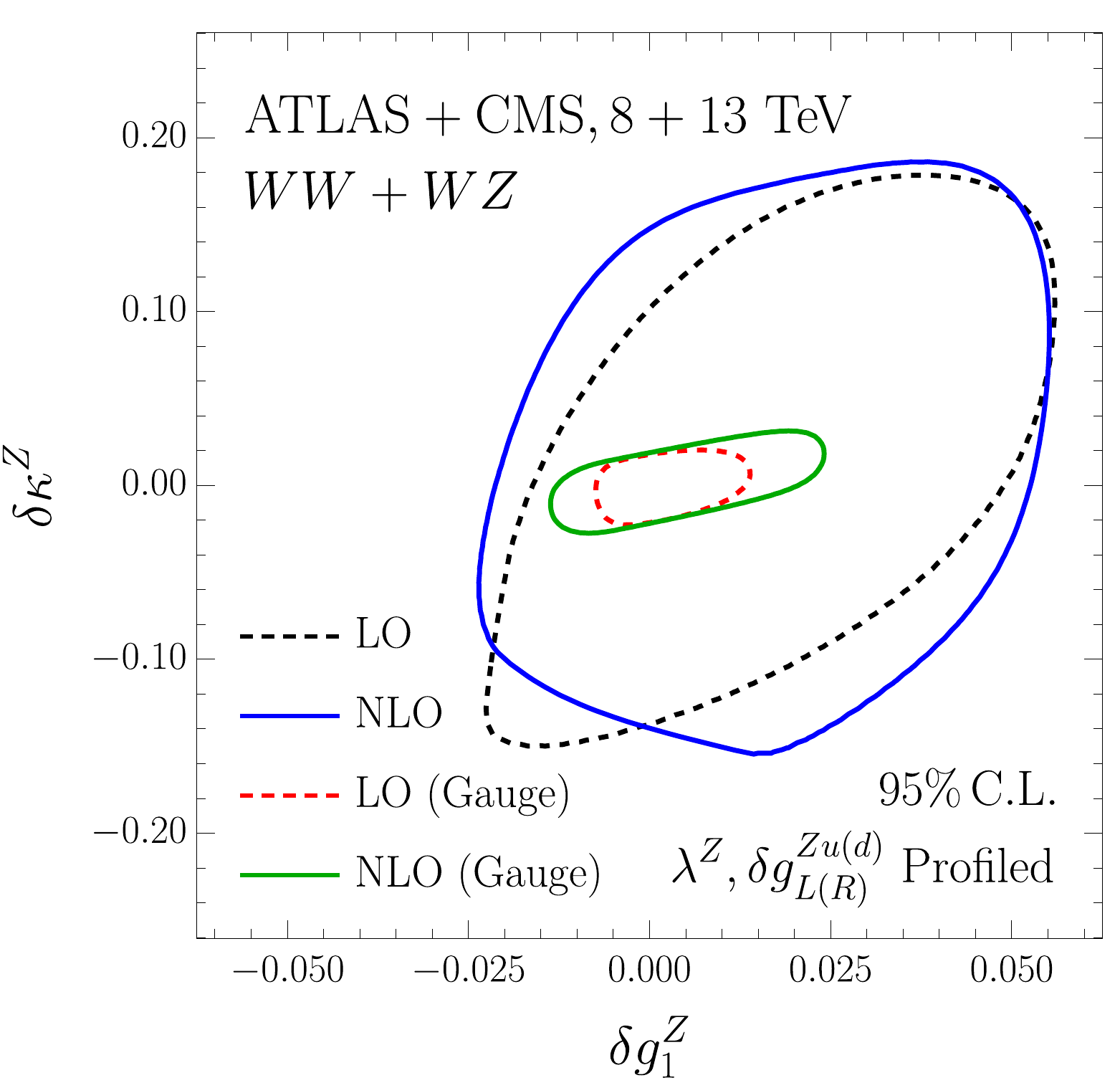}}\qquad~~~\\
\subfigure{\includegraphics[width=0.4\linewidth]{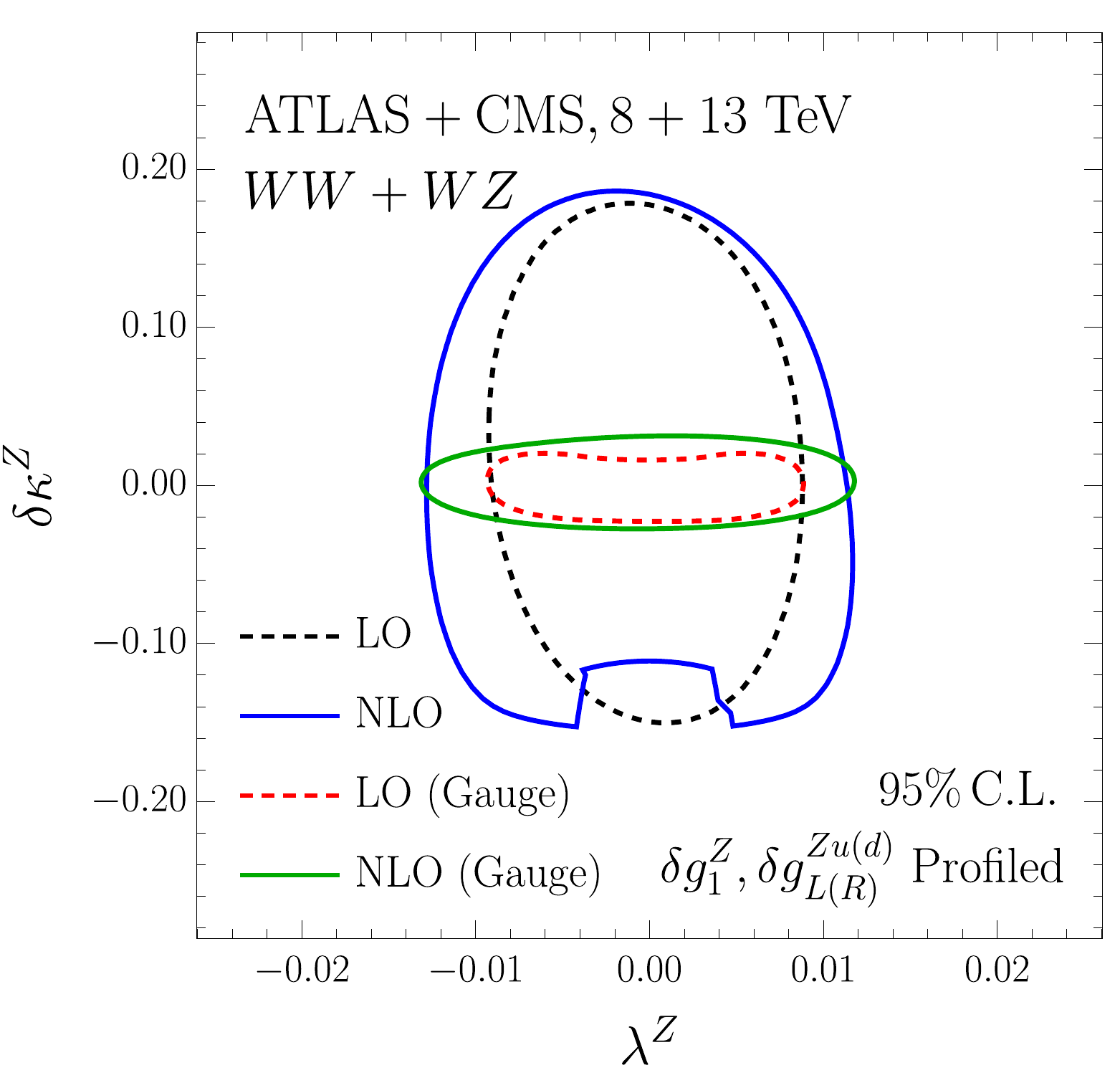}}\qquad~~~
\vskip -0.5cm
\caption{
As in Fig.~\ref{fig:all_proj_contours}, but profiling over the additional operators not shown. The red and green curves show the results profiling only over the gauge couplings at LO and NLO respectively.
}\label{fig:all_prof_contours_gauge}
\end{figure}

\begin{figure}[h]
\centering
\subfigure{\includegraphics[width=0.4\linewidth]{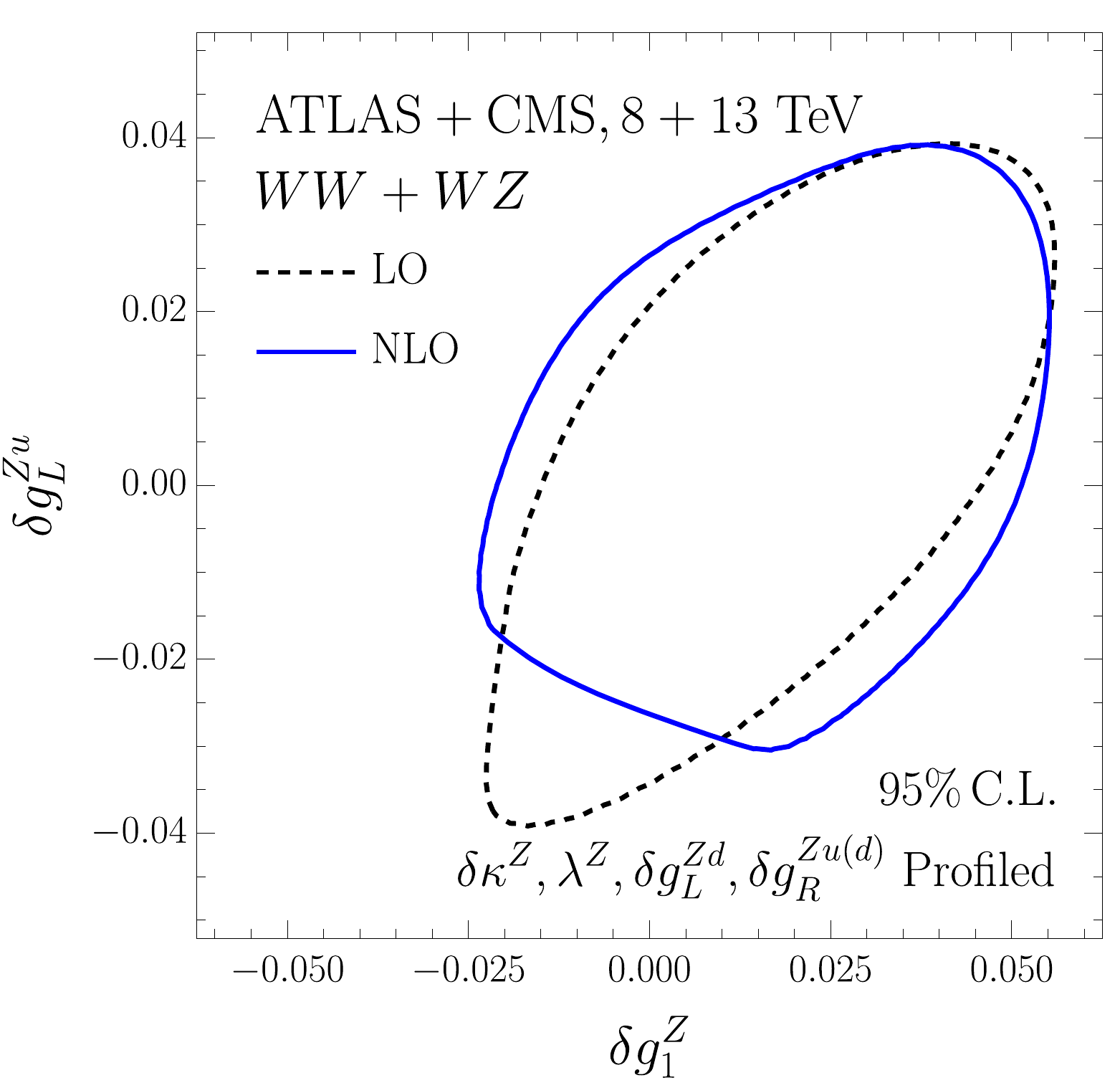}}~~~
\subfigure{\includegraphics[width=0.4\linewidth]{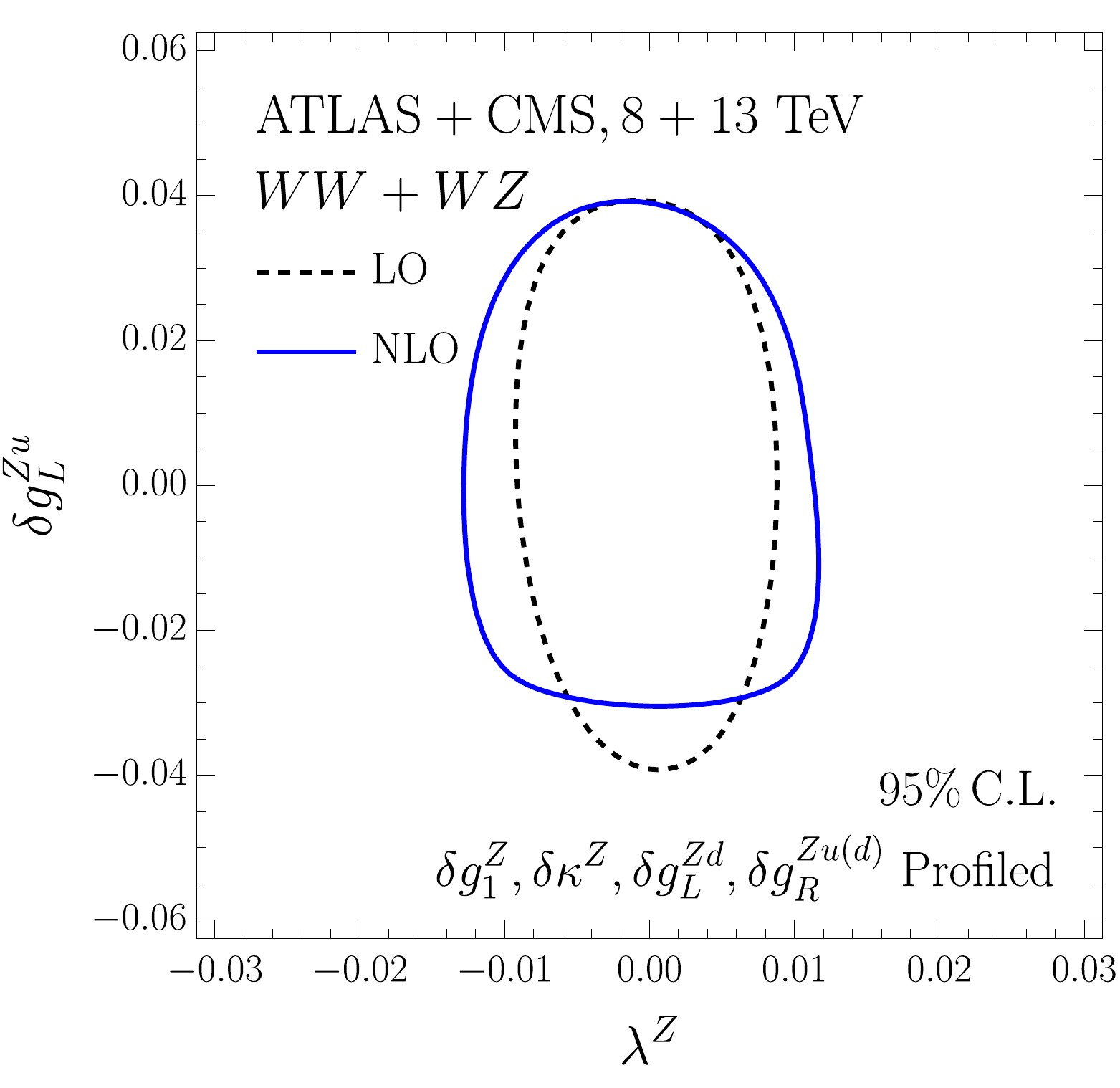}}\qquad~~~\\
\subfigure{\includegraphics[width=0.4\linewidth]{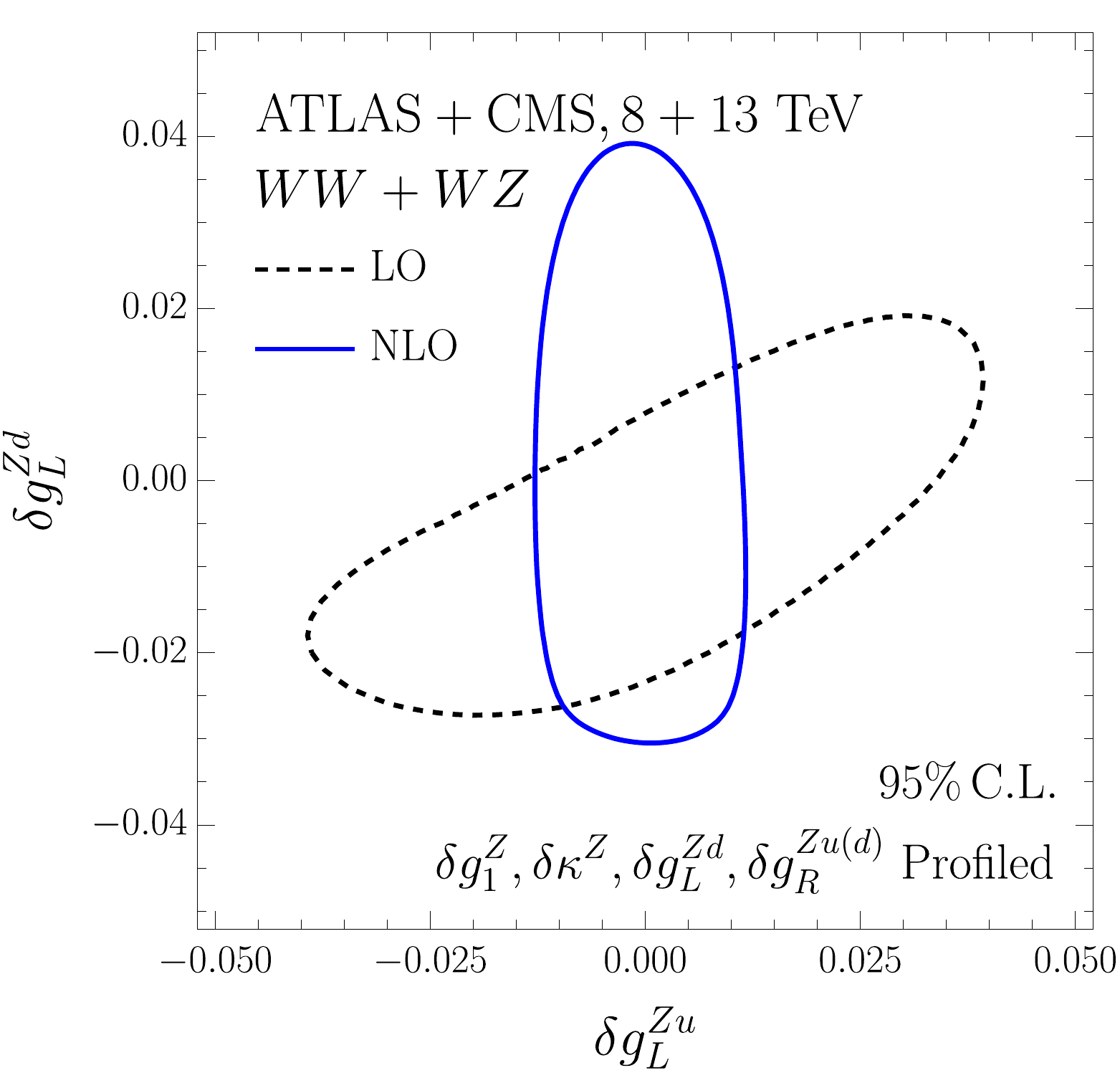}}\qquad~~~
\vskip -0.5cm
\caption{
As in Fig.~\ref{fig:all_proj_contours}, but profiling over the additional operators not shown.}\label{fig:all_prof_contours_ferm}
\end{figure}

We consider the effects of marginalizing over the operators
not shown in each plot. In practice, this is done by minimizing the $\chi^2$ function at each point with respect to the other five couplings. 
In Fig.~\ref{fig:all_prof_contours_gauge}, we show the limits for the three combinations of anomalous gauge couplings, and compare the effects of profiling over the other five  anomalous couplings in black (blue) for LO (NLO) with the results when profiling over only the last gauge coupling in red (green) for LO (NLO). In both cases, the effects of considering the anomalous couplings at NLO weaken the bounds on $\lambda^Z$. The limits on $\delta g_1^Z$ in the $\delta g_1^Z$ -- $\delta\kappa^Z$ plane are also affected, though the effect is more prominent when profiling only over the gauge couplings. 
We also see the result, anticipated in Refs.~\cite{Zhang:2016zsp,Baglio:2017bfe,Butter:2016cvz}, that the limits on the anomalous gauge couplings are generally much weaker when the fermion couplings are allowed to float within their allowed regions. This is only not true in the $\lambda^Z$ direction, as the introduction of $\lambda^Z$ leads to a fundamentally different scaling at high energies for the production of transversely polarized $WZ$ (see Eq.~\ref{eq:wzlims}).

In Fig.~\ref{fig:all_prof_contours_ferm}, we show the constraints in various planes including anomalous fermion couplings, profiling over all five additional parameters. The NLO effects are again apparent, particularly in removing the remaining correlation between $\delta g_L^{Zu}$ and $\delta g_L^{Zd}$.
Finally, we summarize our results in the form of one parameter limits on each of the anomalous couplings considered in Table~\ref{tab:fit_summary}.

\begin{table}
\begin{tabular}{c|cc|cc}
\multirow{2}{*}{~Coupling~} 	& \multicolumn{2}{c|}{LO Allowed Range} 		& \multicolumn{2}{c}{NLO Allowed Range} \\
											& Projected 					& Profiled 						& Projected 					& Profiled \\ \hline
$\delta g_1^Z$ 						& $~[ -0.007,\, 0.005 ]~$	& $~[ -0.015,\, 0.048 ]~$ & $~[ -0.012,\, 0.018 ]~$	& $~[ -0.016,\, 0.049 ]~$ \\
$\lambda^Z$ 						& $~[ -0.008,\, 0.008 ]~$	& $~[ -0.009,\, 0.008 ]~$ & $~[ -0.012,\, 0.010 ]~$	& $~[ -0.012,\, 0.011 ]~$ \\
$\delta \kappa^Z$					& $~[ -0.020,\, 0.015 ]~$	& $~[ -0.135,\, 0.158 ]~$ & $~[ -0.020,\, 0.017 ]~$	& $~[ -0.115,\, 0.168 ]~$ \\
$\delta g_L^{Zu}$					& $~[ -0.002,\, 0.005 ]~$	& $~[ -0.034,\, 0.034 ]~$ & $~[ -0.001,\, 0.008 ]~$	& $~[ -0.022,\, 0.035 ]~$ \\
$\delta g_R^{Zu}$					& $~[ -0.012,\, 0.014 ]~$	& $~[ -0.084,\, 0.086 ]~$ & $~[ -0.012,\, 0.014 ]~$	& $~[ -0.083,\, 0.096 ]~$ \\
$\delta g_L^{Zd}$					& $~[ -0.005,\, 0.002 ]~$	& $~[ -0.025,\, 0.014 ]~$ & $~[ -0.008,\, 0.002 ]~$	& $~[ -0.025,\, 0.023 ]~$ \\
$\delta g_R^{Zd}$					& $~[ -0.018,\, 0.017 ]~$	& $~[ -0.045,\, 0.048 ]~$ & $~[ -0.018,\, 0.017 ]~$	& $~[ -0.053,\, 0.043 ]~$ \\
\end{tabular}
\caption{
$95\%$ C.L. limits on the individual anomalous couplings based on a fit to $W^+W^-$ and $WZ$ data at LO and NLO, projecting out or profiling over the other couplings.
}\label{tab:fit_summary}
\end{table}

\subsection{Validity of our Results}
\begin{figure}[h]
\centering
\subfigure{\includegraphics[width=0.49\linewidth]{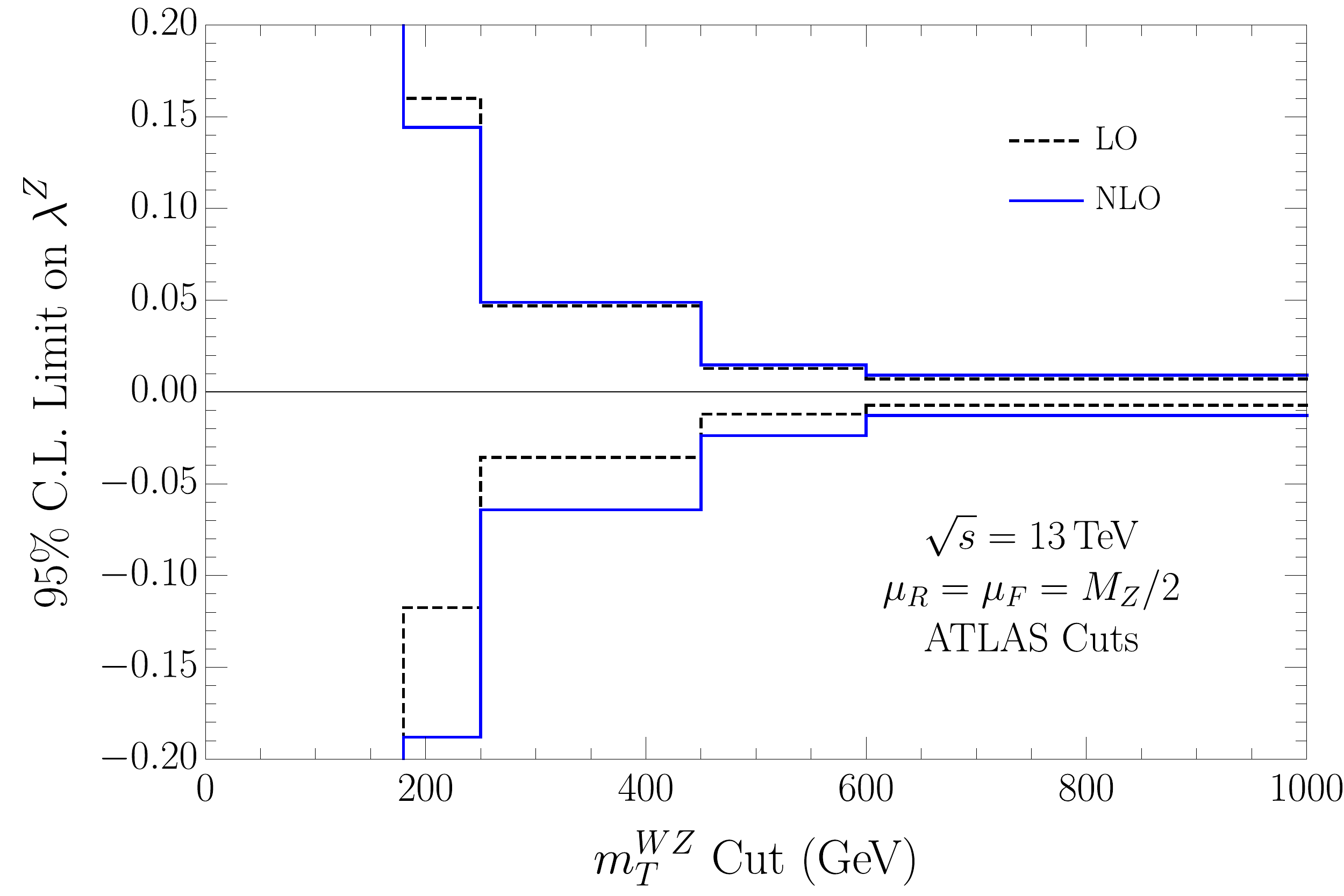}}~~~
\subfigure{\includegraphics[width=0.49\linewidth]{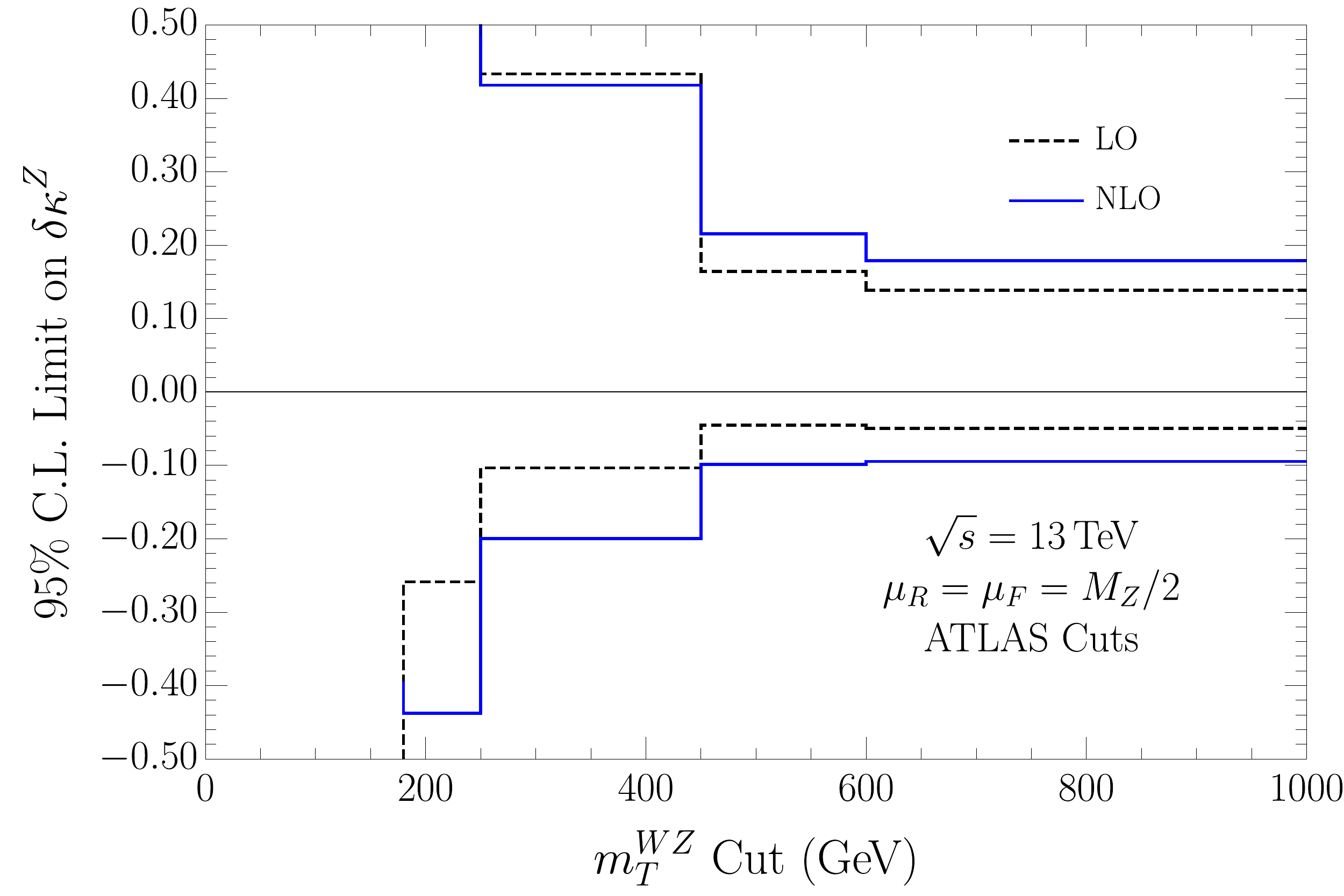}}\qquad~~~\\
\vskip -0.5cm
\subfigure{\includegraphics[width=0.49\linewidth]{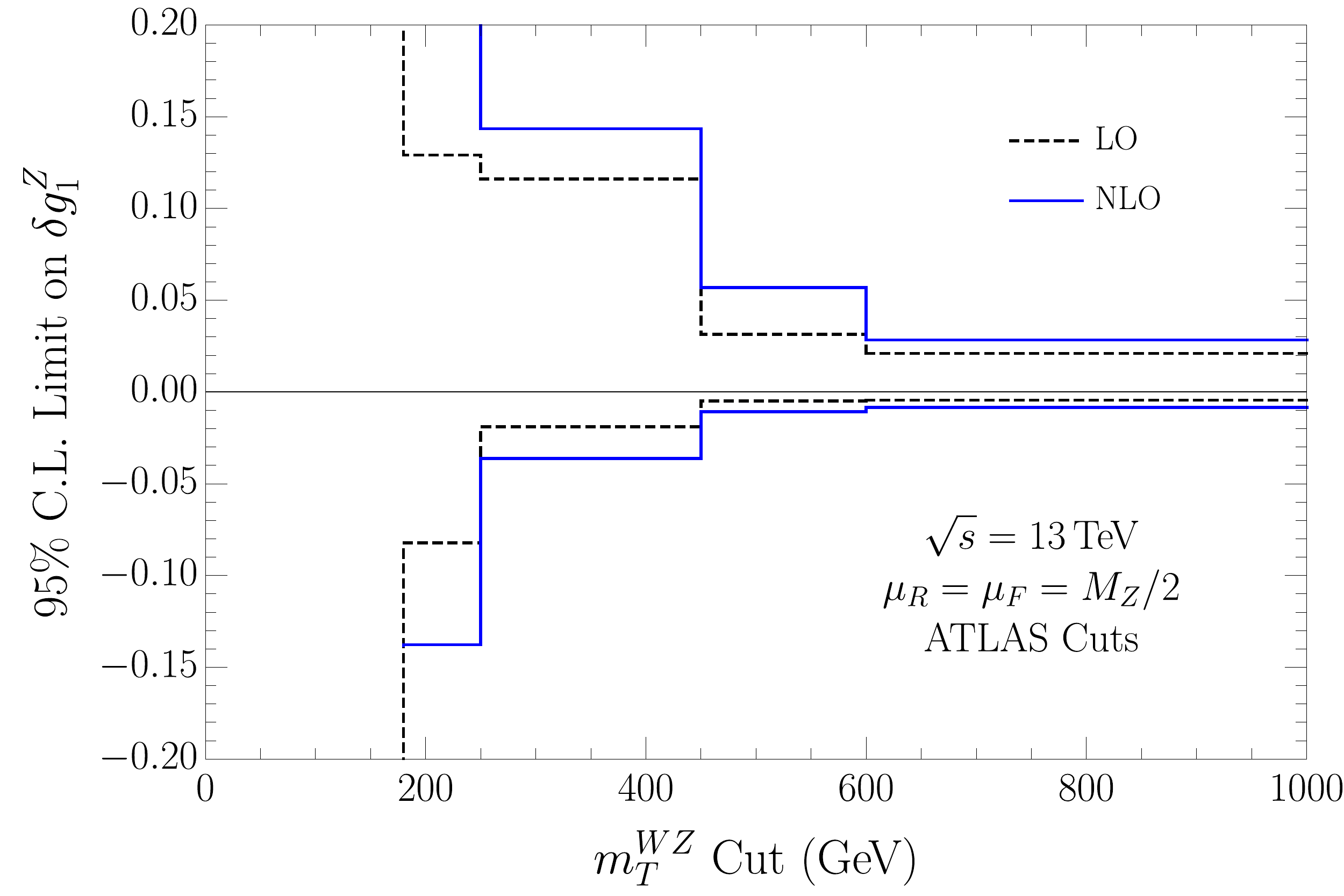}}\qquad~~~
\vskip -0.5cm
\caption{
Single parameter fits to ATLAS $13$ TeV WZ data, when successively removing the high $m_T^{WZ}$ bins.
}
\label{fig:limits}
\end{figure}

The EFT Lagrangian of Eq.~\ref{eq:smeft} is an expansion in powers of (Energy)$^2/\Lambda^2$ and so is only valid for energies
less than the scale $\Lambda$~\cite{Contino:2016jqw,Farina:2016rws}.  We consider in Fig.~\ref{fig:limits} the effects of successively removing the high-energy bins from 
the single parameter fits  to the ATLAS 13 TeV $WZ$ data.  
If we remove the top bin from the fit, $m_T^{WZ}>600$ GeV, then all points in this fit satisfy $m_T ^{WZ}< \Lambda$
 (with $\Lambda=1$ TeV)
 and the EFT is
clearly a valid expansion. In this case, the fit is degraded
by ${\cal{O}}(10\%)$ on $\delta\kappa^Z$ and ${\cal{O}}(30\%)$ for $\lambda^Z$ and $\delta g_1^Z$, and the NLO QCD effects remain important.  It would be
interesting to have experimental results where the overflow
bin is explicitly separated, so that the maximum energy of the data points in the last bin is clear.
The result that interesting limits can be obtained even disregarding the highest-energy bin has been demonstrated
using machine learning
for the case of $WH$ production in Ref. \cite{Brehmer:2019gmn}.

\section{Conclusions}\label{sec:conc}
The SMEFT NLO QCD calculation for  $pp\rightarrow WZ\rightarrow (l^\prime \nu^\prime) (l^+l^-)$
 has been included  in the {\tt POWHEG-BOX} and 
the primitive cross sections needed to reproduce our results at  $8$
and $13$\,TeV can be found 
at
\url{https://quark.phy.bnl.gov/Digital_Data_Archive/dawson/wz_19}.
The NLO QCD effects are significant for $WZ$ production and have an
important effect on the global fits to anomalous couplings. 
The $\mathcal{O}(1/\Lambda^4)$ terms dominate over the
$\mathcal{O}(1/\Lambda^2)$ terms also when NLO QCD corrections are
taken into account.
We emphasize again that these results should be interpreted as only a
first step in a profiled, global analysis, as all of the couplings ---
especially the fermionic ones --- will be further constrained, and
some flat directions removed, by data from LEP and Higgs
measurements.

\begin{acknowledgments}
We thank Anke Biek\"otter and Tilman Plehn for useful discussions of
the global fits and Ian Lewis for insights into gauge boson pair
production. SD is supported by the United States Department of Energy
under Grant Contract DE-SC0012704 and is grateful to the University of
T\"ubingen, where this work was started. The work of SH was supported
in part by the National Science Foundation grant PHY-1620628 and in
part by grant PHY-1915093. SH was also supported by the
U.S. Department of Energy, Office of Science, Office of Workforce
Development for Teachers and Scientists, Office of Science Graduate
Student Research (SCGSR) program. The SCGSR program is administered by
the Oak Ridge Institute for Science and Education (ORISE) for the
DOE. ORISE is managed by ORAU under contract number
DE-SC0014664. J.B. acknowledges the support from the Carl-Zeiss
foundation. Parts of this work were performed thanks to the support of
the State of Baden-W\"urttemberg through bwHPC and the DFG through the
grant no. INST 39/963-1 FUGG.
\end{acknowledgments}

\appendix
\section{Fits to $W^+W^-$ Production}\label{app:ww_limits}

\begin{figure}
\centering
\subfigure{\includegraphics[width=0.4\linewidth]{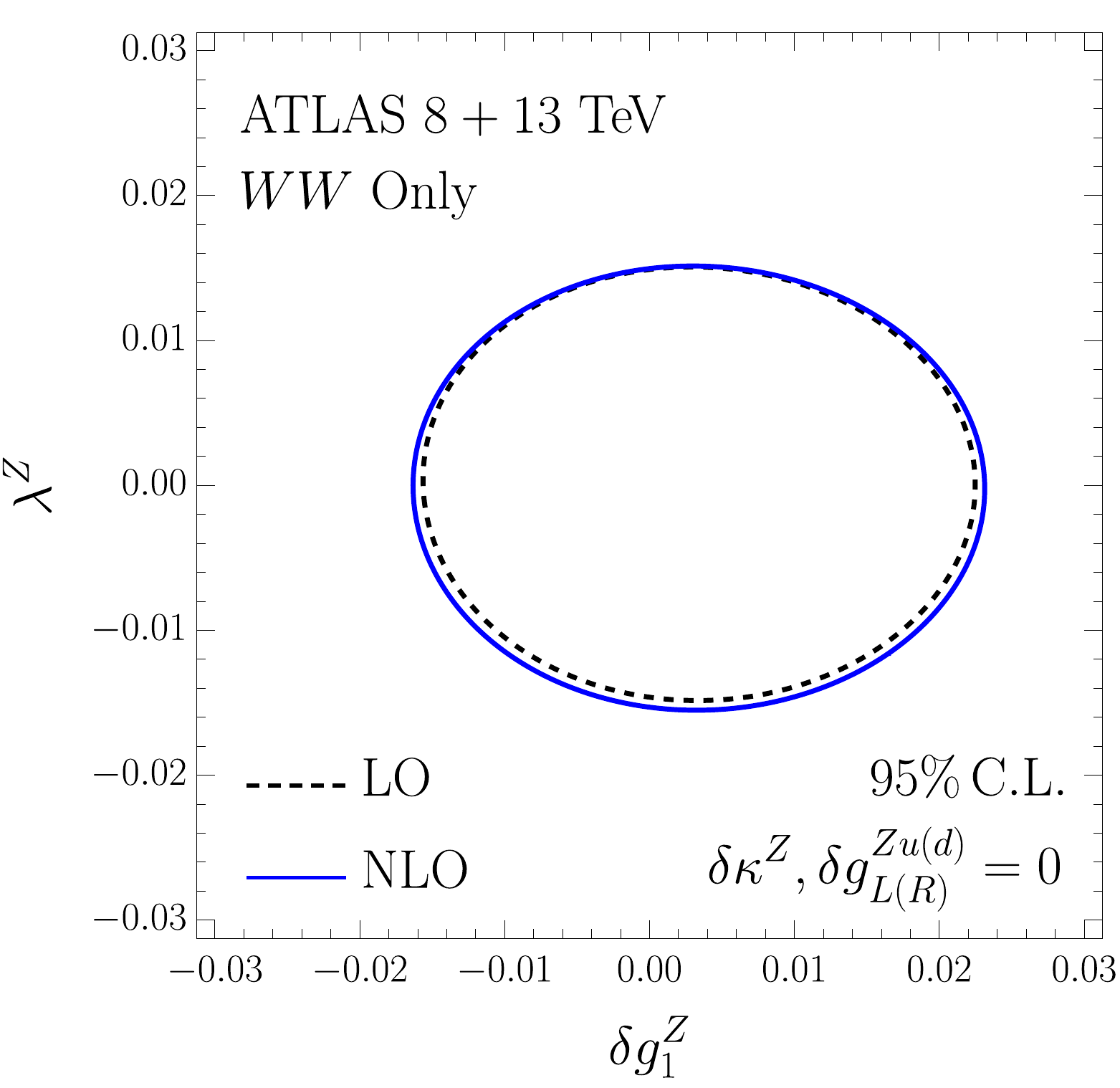}}~~~
\subfigure{\includegraphics[width=0.4\linewidth]{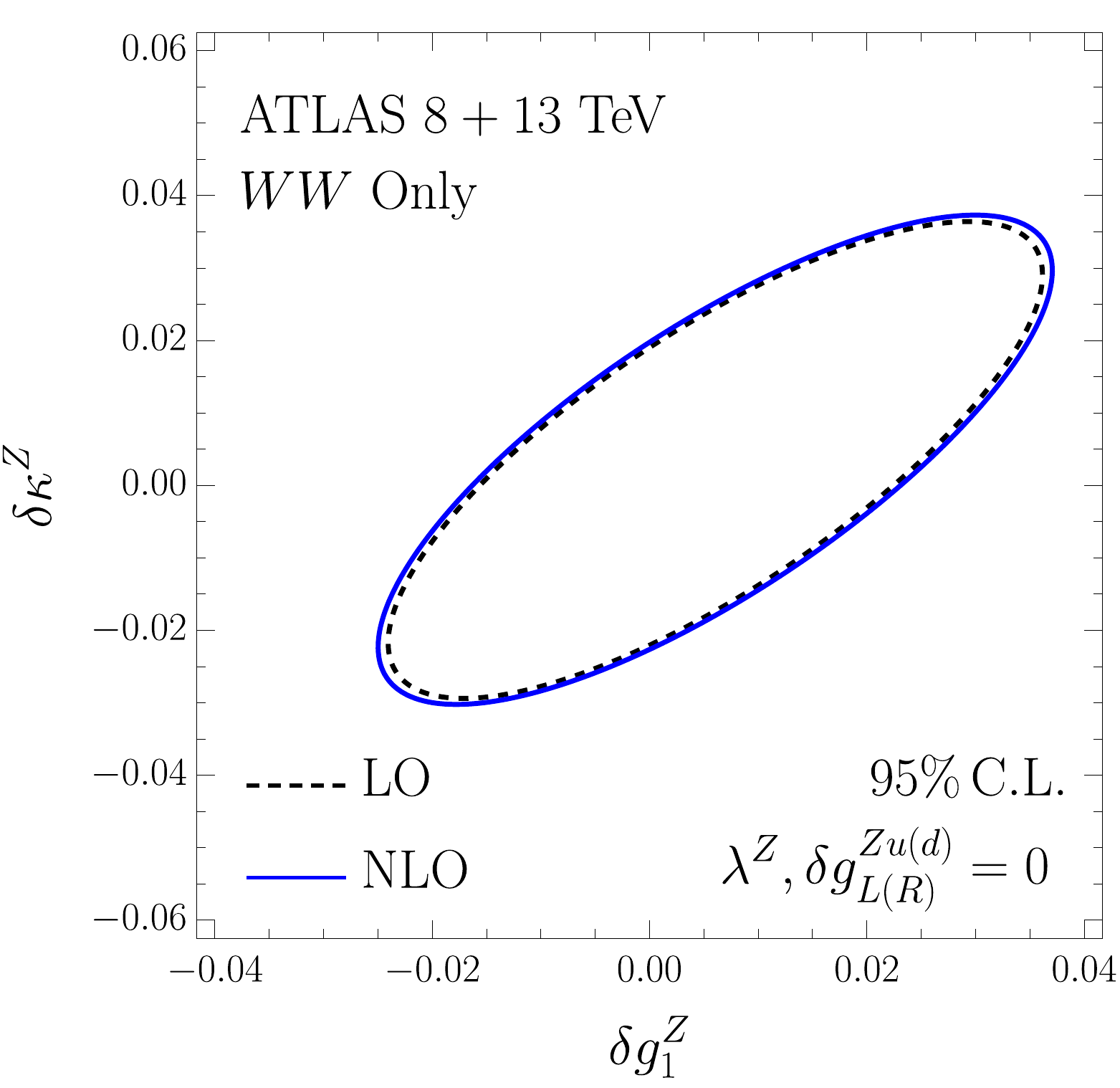}}\qquad~~~\\
\subfigure{\includegraphics[width=0.4\linewidth]{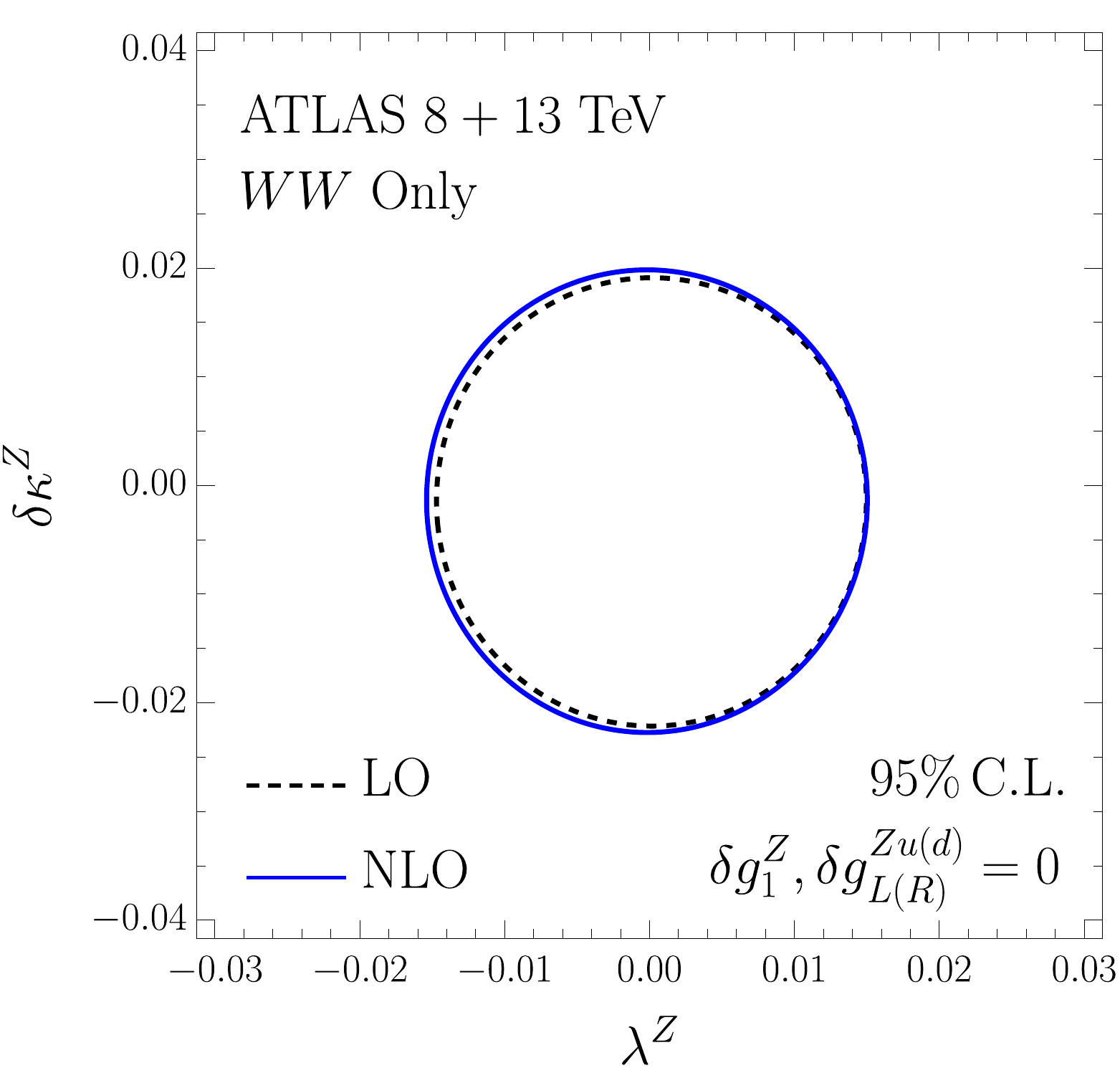}}~~~
\subfigure{\includegraphics[width=0.4\linewidth]{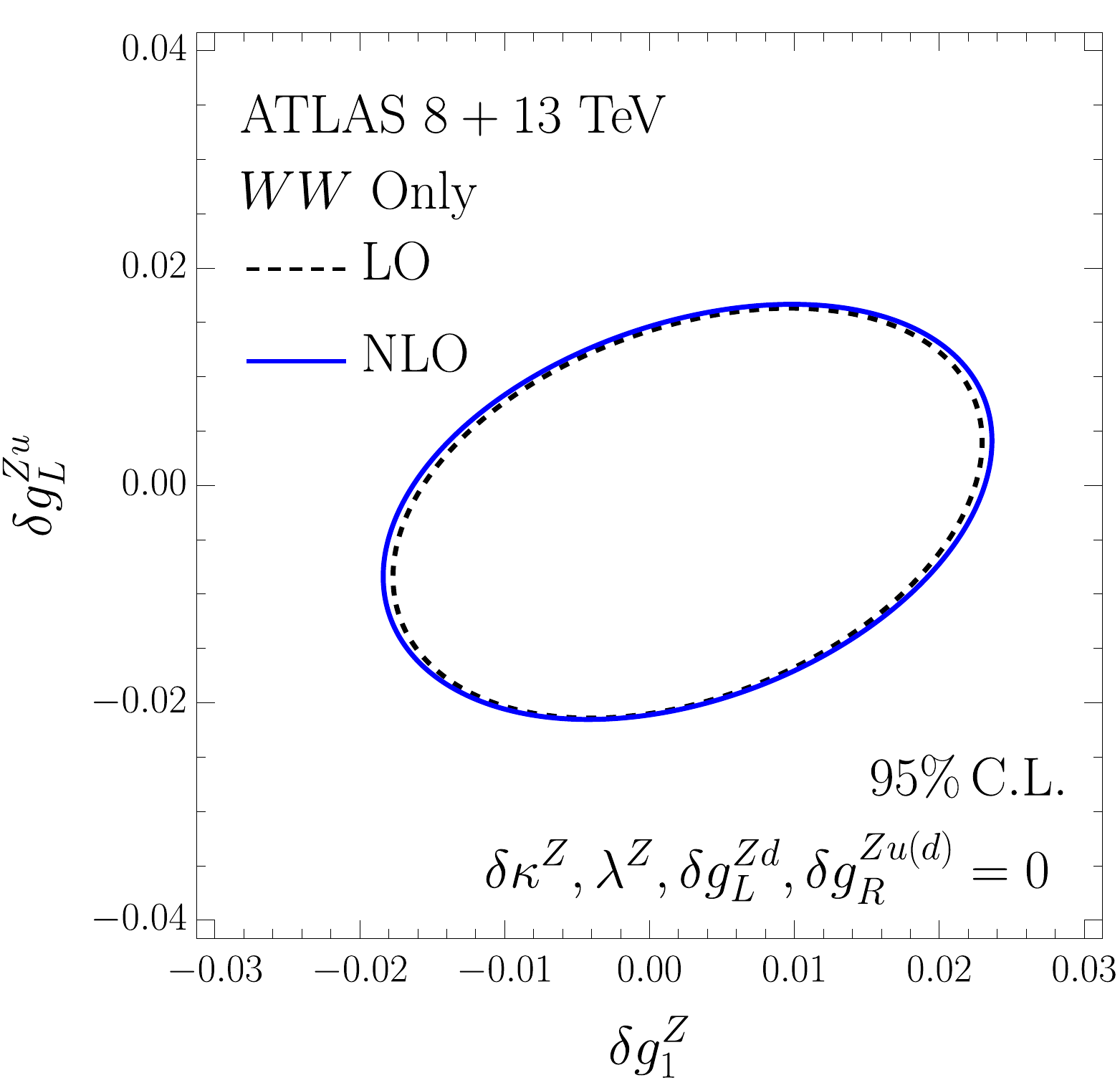}}\qquad~~~\\
\subfigure{\includegraphics[width=0.4\linewidth]{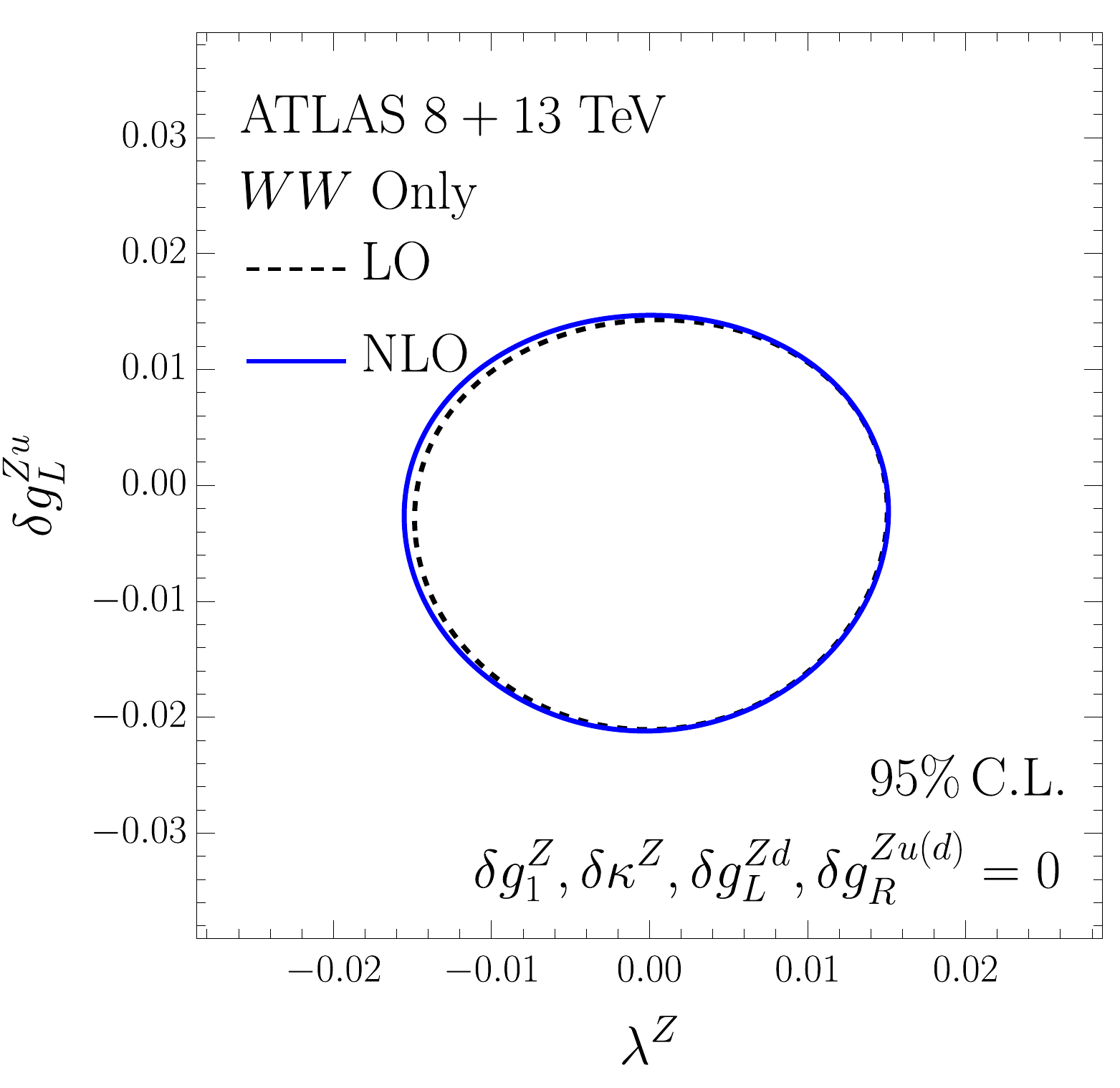}}~~~
\subfigure{\includegraphics[width=0.4\linewidth]{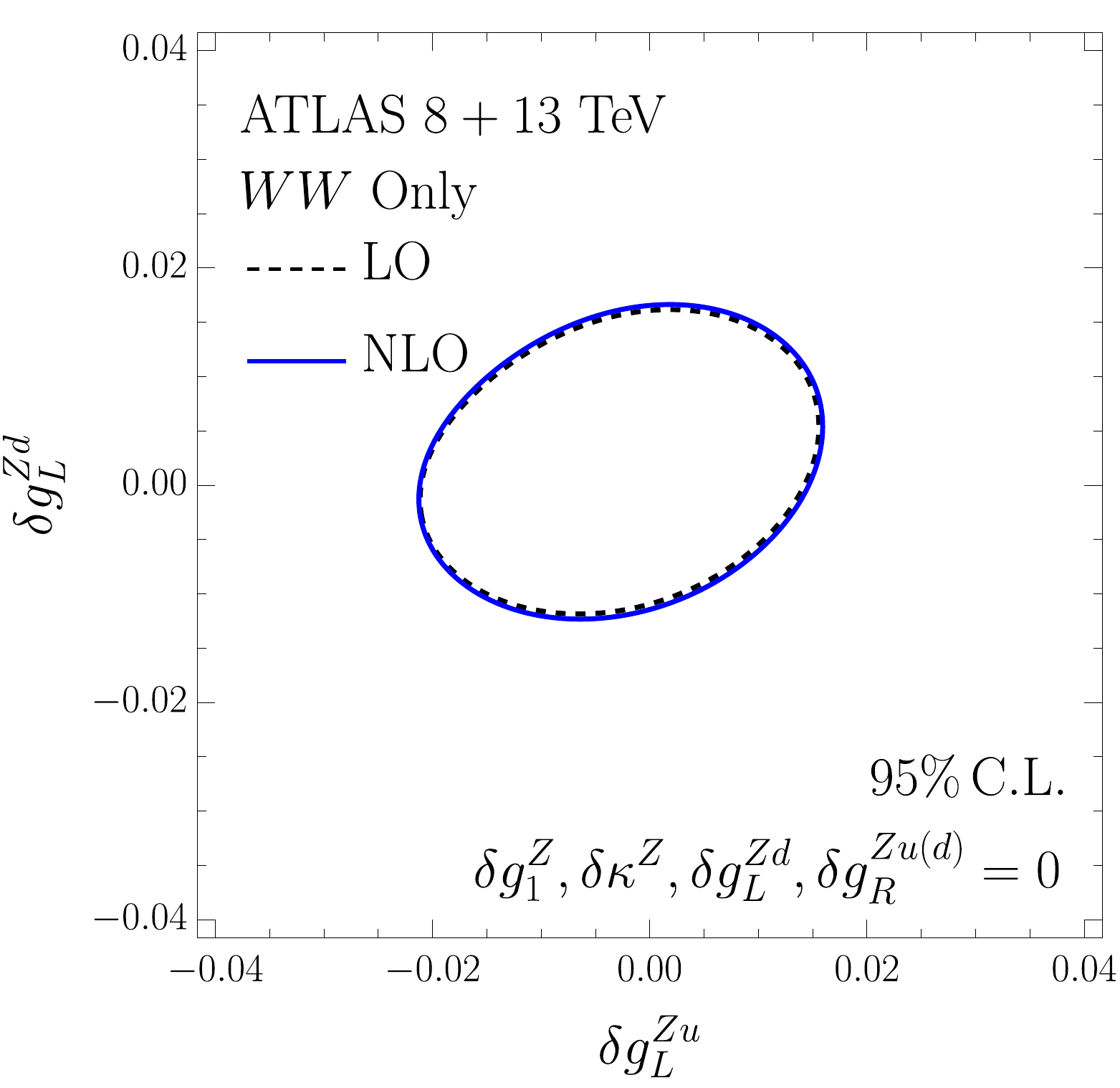}}\qquad~~~
\vskip -0.5cm
\caption{
As in Fig.~\ref{fig:wz_proj_contours}, but using only $WW$ data.
}\label{fig:ww_proj_contours}
\end{figure}

In this Appendix we present constraints on the anomalous gauge and fermion couplings based on only the $W^+W^-$ data from ATLAS detailed in Table~\ref{tab:data}. In Fig.~\ref{fig:ww_proj_contours}, we show the two dimensional limits setting the other anomalous couplings to zero.
It is apparent that the NLO QCD effects do not have a significant impact on fits to the $W^+W^-$ data alone.

\clearpage

\section{Truncation to $1/\Lambda^2$}\label{app:truncation}

\begin{figure}
\centering
\subfigure{\includegraphics[width=0.42\linewidth]{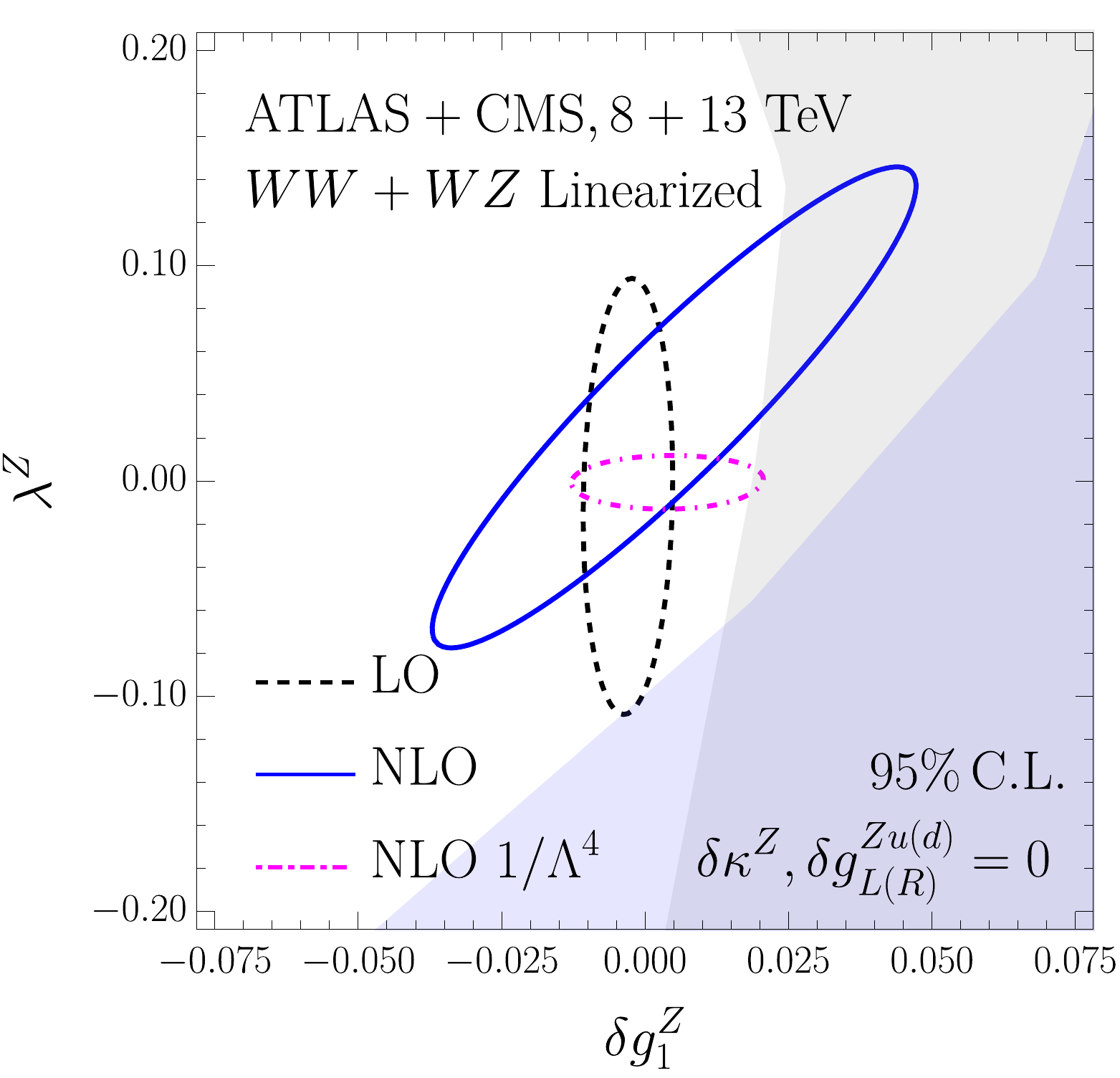}}~~~
\subfigure{\includegraphics[width=0.42\linewidth]{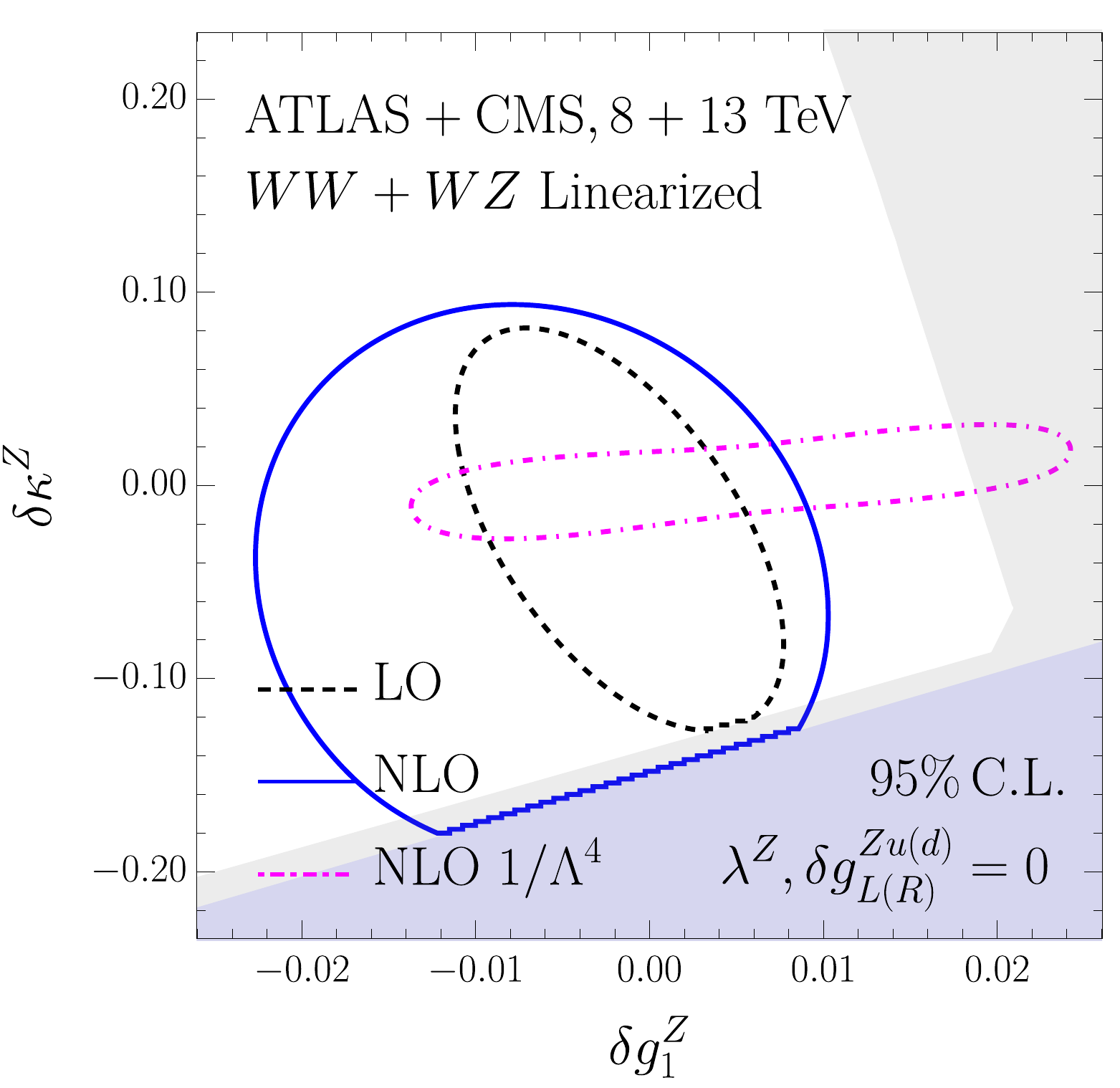}}\qquad~~~\\
\subfigure{\includegraphics[width=0.42\linewidth]{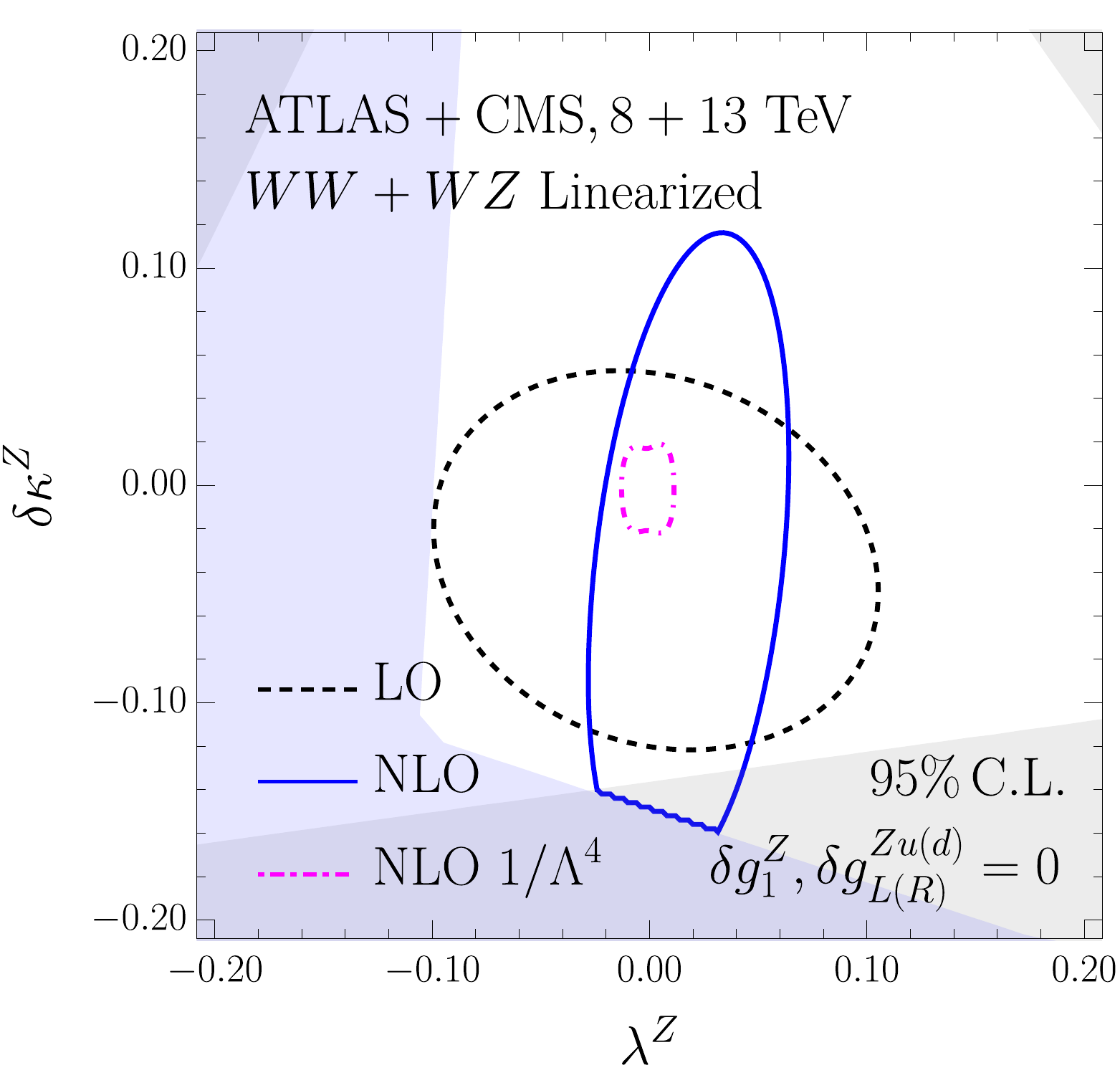}}~~~
\subfigure{\includegraphics[width=0.42\linewidth]{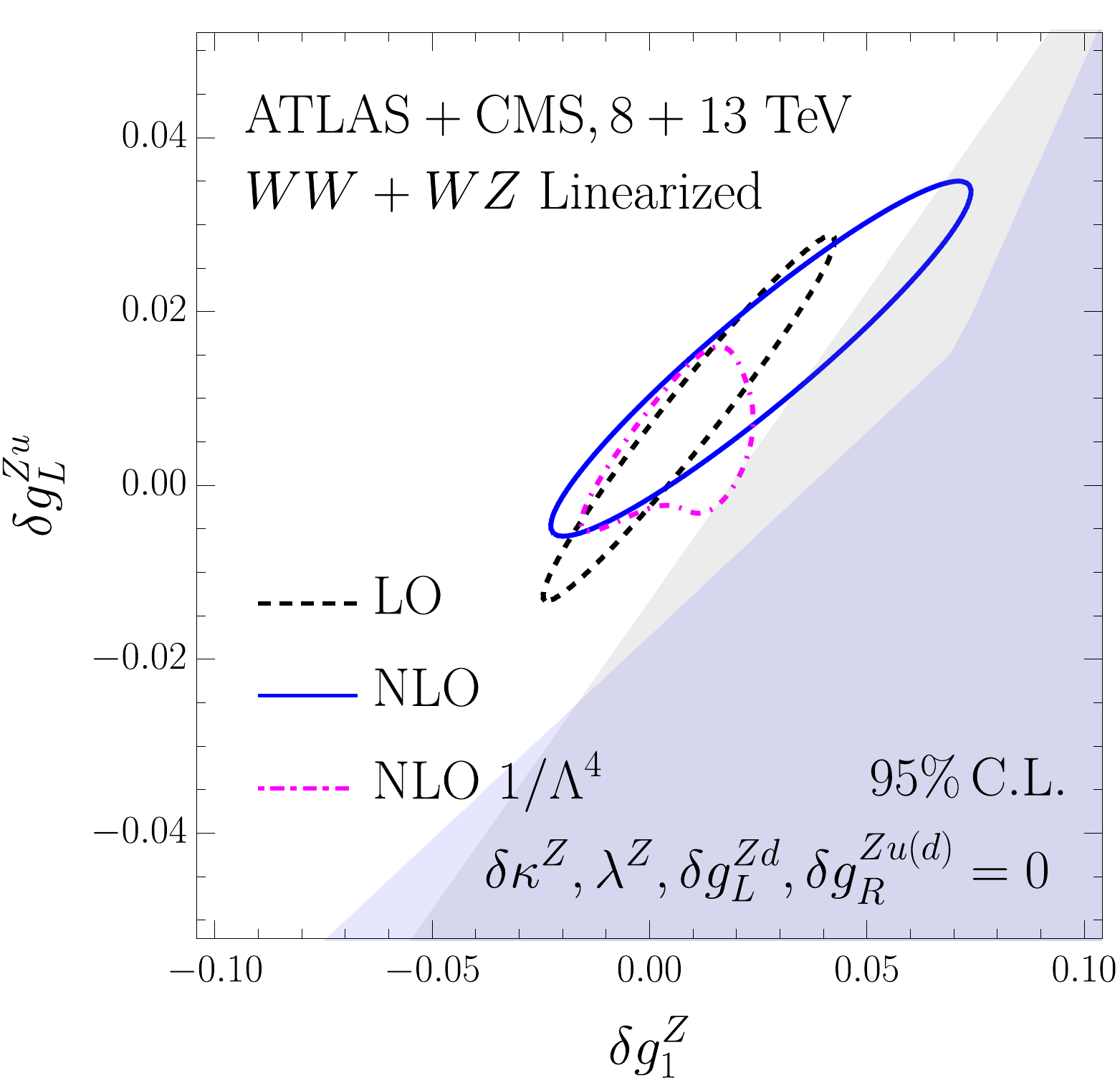}}\qquad~~~\\
\subfigure{\includegraphics[width=0.42\linewidth]{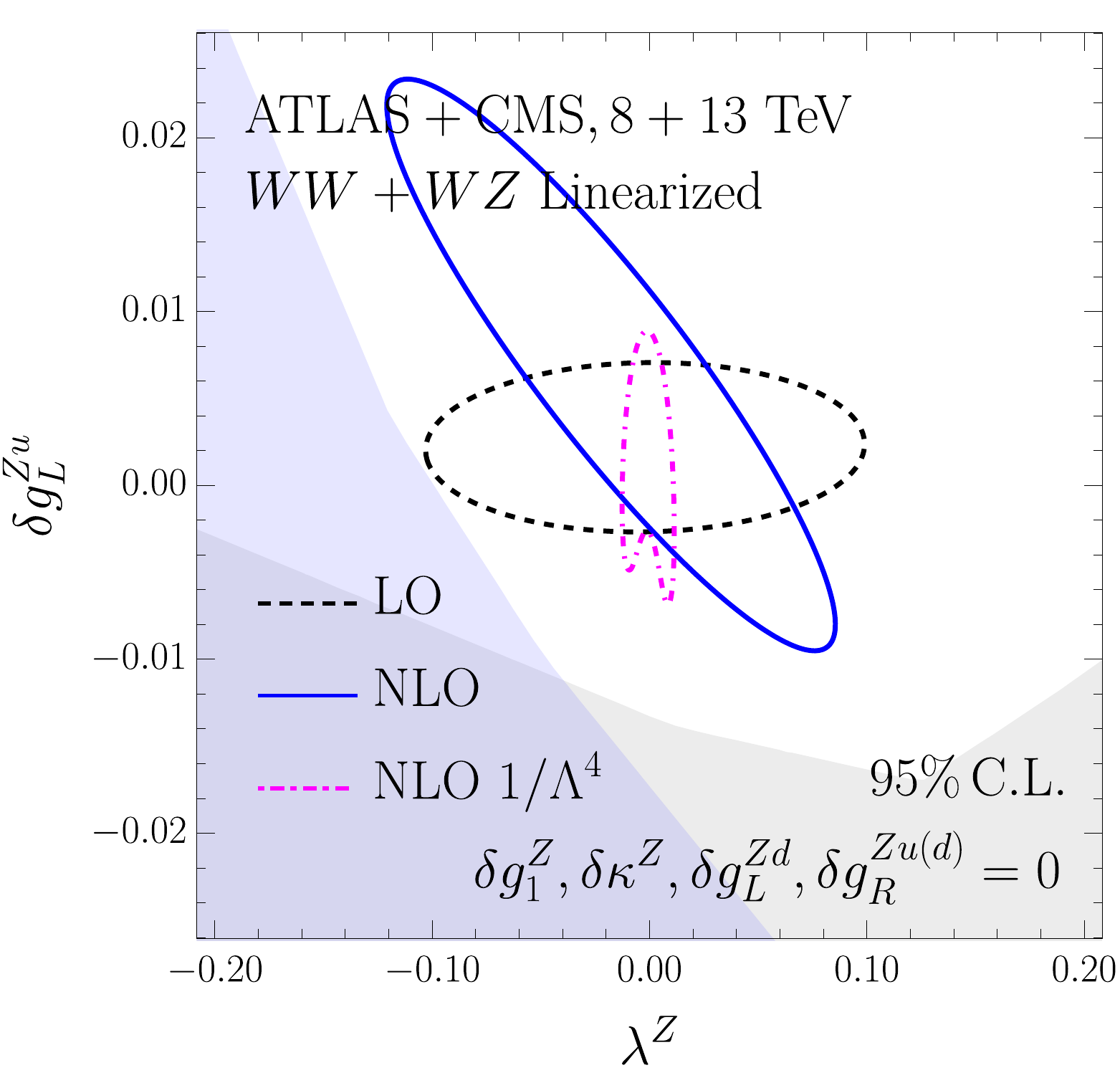}}~~~
\subfigure{\includegraphics[width=0.42\linewidth]{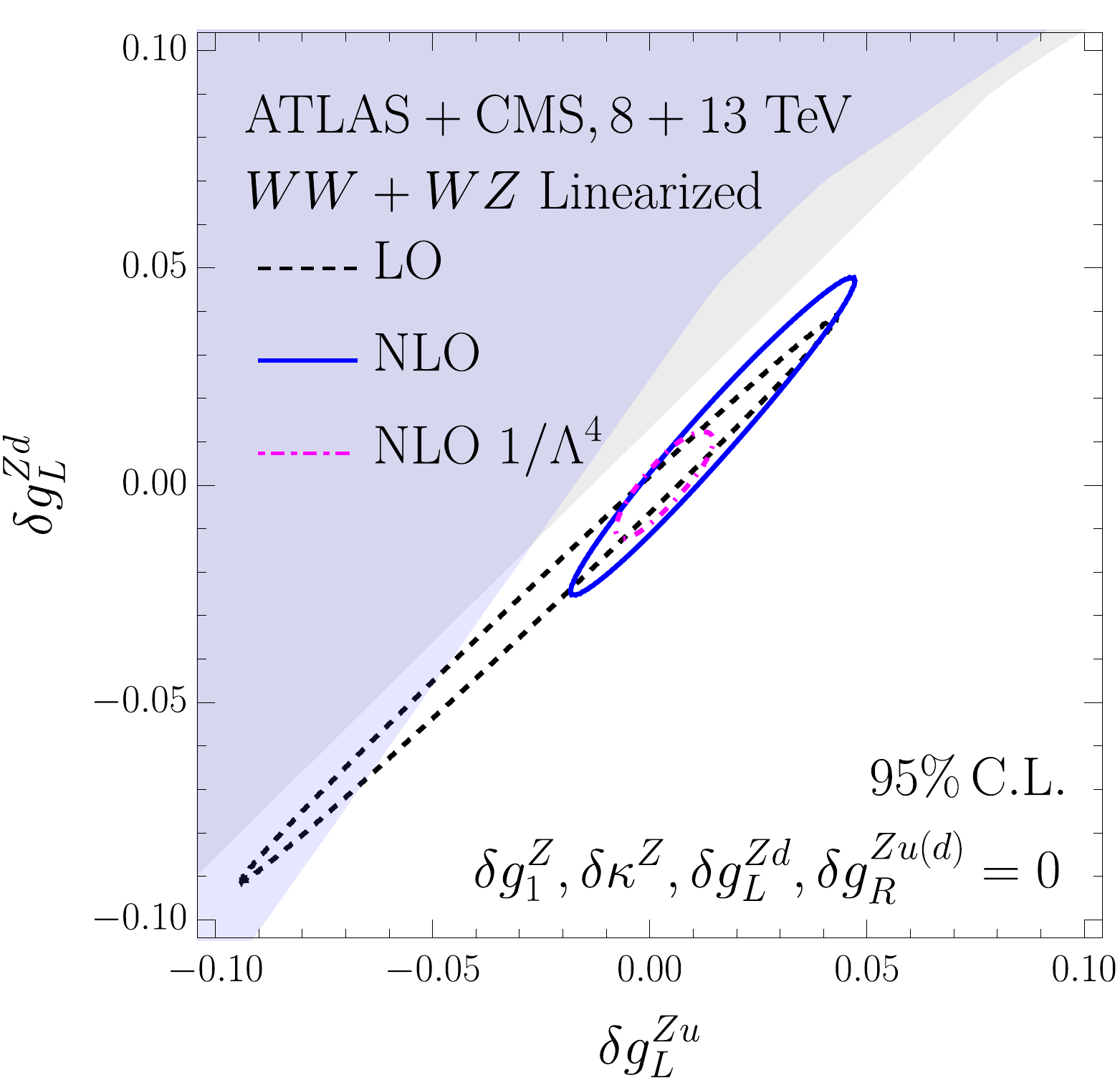}}\qquad~~~
\vskip -0.5cm
\caption{
The same as Fig.~\ref{fig:all_proj_contours}, but keeping only the leading terms in $1/\Lambda^2$. Regions where the predicted cross section in any of the bins used for the fit becomes negative are shown in grey (blue for NLO). We include the $1/\Lambda^4$ NLO results (pink)
for comparison.}
\label{fig:all_lin_contours}
\end{figure}

If the anomalous couplings are assumed to be small, then the dominant contribution is from the ${\cal{O}}({1/\Lambda^2})$ 
terms which are linear in the anomalous couplings.  The results of such a linearized fit are shown in
Fig.~\ref{fig:all_lin_contours}.
Since this fit does not include the full amplitude-squared, but just the interference terms it is not guaranteed to be positive definite. The regions with negative cross sections are shaded in grey (blue) for the LO (NLO) predictions, corresponding to regions where the linear approximation is not valid.  A comparison of
the linear and quadratic fits of Fig.~\ref{fig:all_lin_contours} demonstrates that the fits are dominated by the quadratic ${\cal{O}}({1/\Lambda^4})$  contributions, and hence the data
is not yet sensitive to weak anomalous couplings where the linear approximation would be valid.

\newpage
\bibliographystyle{utphys}
\bibliography{ww}

\end{document}